\documentclass{ws-rv975x65}
\usepackage[square]{ws-rv-van}

\usepackage{amsmath}
\usepackage{amssymb}
\usepackage{gastex}
\usepackage[eurosym]{eurofont}

\def\cqfd{\skip10=\parfillskip\parfillskip=0pt\enspace
\hfill\symbolecqfd\par\parfillskip=\skip10\par\medskip}
\def\symbolecqfd{\rlap{$\sqcap$}$\sqcup$}
\def\eop{\end{proof}}

\newenvironment{sketchofproof}{\rm \trivlist \item[\hskip \labelsep{\bf\it
Sketch of proof.}]}{\cqfd\endtrivlist}
\def\sketch{\begin{sketchofproof}}
\def\eosketch{\end{sketchofproof}}

\def\calA{\mathcal{A}}
\def\calB{\mathcal{B}}
\def\calC{\mathcal{C}}

\def\O{\mathcal{O}}
\def\calP{\mathcal{P}}
\def\calR{\mathrel{\mathcal{R}}}

\let\phi\varphi
\let\epsilon\varepsilon
\def\inv{^{-1}}

\def\synt(#1){\textsf{Synt}(#1)}

\font\petite=cmmi10 at 8pt
\def\malcev{\mathbin{\hbox{$\bigcirc$\rlap{\kern-9pt\raise0,75pt\hbox{\petite m}}}}}

\def\FO{\textsf{FO}}
\def\MSO{\textsf{MSO}}

\def\rat{\textsf{Rat}}
\def\extrat{\textsf{X-Rat}}
\def\automate{(Q,T,I,F)}
\let\auto=\automate
\def\autop{(Q',T',I',F')}
\def\calAem{{\calA}_{\rm trim}}

\def\et{\land}
\def\ou{\lor}
\def\vrai{\textbf{true}}
\def\Dom{\textsf{Dom}}
\newcommand{\trans}[1]{\stackrel{#1}{\longrightarrow}}
\def\FG{{\rm Pref}}
\def\FD{{\rm Suff}}
\def\F{{\rm Fact}}
\newcommand{\SM}{\mathop{\textrm{SW}}}
\newcommand{\shuf}{\mathbin{\sqcup \! \sqcup}}

\makeindex

\begin{document}

\setcounter{page}{3}

\chapter[An Introduction to Finite Automata]{An Introduction to
  Finite Automata and their\\ Connection to Logic}
\label{chap:weil}
\author{Howard Straubing\footnote{Work partially supported by NSF Grant CCF-0915065}}

\address{Computer Science Department, Boston College,\\ Chestnut Hill, Massachusetts, USA}

\author[H. Straubing and P. Weil]{Pascal Weil\footnote{Work partially supported by ANR 2010 BLAN 0202 01 FREC}}

\address{LaBRI, Universit\'e de Bordeaux and CNRS, Bordeaux, France}


\begin{abstract}
This introductory chapter is a tutorial on finite automata.  We present the standard material on determinization and minimization, as well as an account of the equivalence of finite automata and monadic second-order logic.  We conclude with an introduction to the syntactic monoid, and as an application give a proof of the equivalence of first-order definability and aperiodicity.
\end{abstract}

\section{Introduction}

\subsection{Motivation}

{The word {\it automaton} (plural: {\it automata}) was originally used to refer to devices like clocks and watches, as well as mechanical marvels built to resemble moving humans and animals, whose internal  mechanisms are hidden and which thus appear to operate spontaneously. In theoretical computer science, 
the {\it finite automaton} is among the simplest models of computation:  A device that can be in one of finitely many {\it states,} and that receives a discrete sequence of inputs from the outside world, changing its state accordingly.  This is in marked contrast to more general and powerful models of computation, such as Turing machines, in which the set of global states of the device---the so-called {\it instantaneous descriptions}---is infinite.  A finite automaton is more akin to the control unit of the Turing machine (or, for that matter, the control unit of a modern computer processor), in which the present state of the unit and the input symbol under the reading head determine the next state of the unit, as well as signals to move the reading head left or right and to write a symbol on the machine's tape. The crucial distinction is that while the Turing machine can record and consult its entire computation history, all the\hfilneg} 

\newpage

\noindent
information that a finite automaton can use about the sequence of inputs it has seen is represented in its current state.

\enlargethispage{13pt}

But as rudimentary as this computational model may appear, it has a rich theory, and many applications.  In this introductory chapter, we will present the core theory: that of a finite automaton reading a finite {\it word,} that is, a finite string of inputs, and using the resulting state to decide whether to accept or reject the word.  The central question motivating our presentation is to determine what properties of words can be decided by finite automata. Subsequent chapters will present both generalizations of the basic model (to devices that read infinite words, labeled trees, {\it etc.}) and to applications.  An important theme in this chapter, as well as throughout the volume, is the close connection between automata and formal logic.

\subsection{Plan of the chapter}

In Section~\ref{section:automata}, we introduce finite automata as devices for recognizing formal languages, and show the equivalence of several  variants of the basic model, most notably the equivalence of deterministic and nondeterministic automata. Section~\ref{section:logic} describes B\"uchi's sequential calculus, the framework in predicate logic for describing properties of words that are recognizable by finite automata.  In Section~\ref{section:kleene-buchi} we prove what might well be described as the two fundamental theorems of finite automata:  that the languages recognized by finite automata are exactly those definable by sentences of the sequential calculus, and also exactly those definable by rational expressions (also called regular expressions).  Section~\ref{section:pumping} presents methods that can be used to show certain languages cannot be recognized by finite automata.  The last sections, \ref{section:minimal} and \ref{sec: FO SF Ap}, have a more algebraic flavor:  we introduce both the minimal automaton and the syntactic monoid of a language, and prove the important McNaughton-Sch\"utzenberger theorem describing the languages definable in the first-order fragment of the sequential calculus.

\subsection{Notation} 

Throughout this chapter, $A$ denotes a finite \emph{alphabet}\index{alphabet}, that is, a finite non-empty set.  Elements of $A$ are called \emph{letters}\index{letter}, and a finite sequence of letters is called a \emph{word}\index{word}.  We denote words simply by concatenating the letters, so, for example, if $A=\{a,b,c\}$, then $aabacba$ is a word over $A$.  The \emph{empty sequence}\index{word!empty} is considered a word, and we use $\epsilon$ to denote this sequence.  The set of all words over $A$ is denoted $A^*$, and the set of all nonempty words is denoted $A^+$.  The \emph{length}\index{length} of the word $w$, that is, the number of letters in $w$, is denoted $|w|$.

If $u,v\in A^*$ then we can form a new word $uv$ by concatenating the two sequences. Concatenation of words is obviously an associative and (unless $A$ has a single element) noncommutative operation on $A^*$.  We have
\begin{align*}
|uv| &= |u|+|v|,\textrm{ and}\\
u\,\epsilon &= \epsilon\,u = u.
\end{align*}
(Other texts frequently use $\Lambda$ or $1$ to denote the empty word. The latter choice is justified by the second equation above.)

A subset of $A^*$ is called a \emph{language}\index{language} over $A$.







\subsection{Historical note and references}

This chapter contains a modern presentation of material that goes back more than fifty years.  The reader can find other accounts in classic papers and texts:  The equivalence of finite automata and rational expressions given in Section~\ref{section:kleene-buchi} was first described by Kleene in ~\cite{Kleene1956}.  The connection with monadic second-order logic was found independently by Trakhtenbrot~\cite{Trakhtenbrot:1958} and B\"uchi~\cite{Buchi:1960}.  

Nondeterministic automata were introduced by Rabin and Scott~\cite{RabinScott1959ibm}, who showed their equivalence to deterministic automata. Minimization of finite-state devices (framed in the language of switching circuits built from relays) is due to Huffman~\cite{Huffman:1954}.  The simple congruential account of minimization that we give originates with Myhill~\cite{Myhill1957} and Nerode~\cite{Nerode:1958}.
 
 The equivalence of aperiodicity of the syntactic monoid with star-freeness is due to Sch\"utzenberger~\cite{Schutzenberger:1965}, and the connection with first-order logic is from McNaughton and Papert~\cite{McNaughton&Papert:1971}.  Our account of these results relies heavily on an argument given in Wilke~\cite{Wilke:1999}.
 
Rational expressions, determinization and minimization have become part of the basic course of study in theoretical computer science, and as such are described in a number of undergraduate textbooks.  Hopcroft and Ullman~\cite{HopcroftUllman1979book}, Lewis and Papadimitriou~\cite{LewisPapadimitriou:1981} and the more recent Sipser~\cite{Sipser:2006} are notable examples.  A more technical and algebraically-oriented account is given in the monograph by Eilenberg~\cite{Eilenberg:1974, Eilenberg:1976}. An algebraic view of automata is developed by Sakarovitch~\cite{Sakarovitch:2009}. Detailed accounts of the connection between automata, logic and algebra can be found in Straubing~\cite{Straubing:1994} and Thomas~\cite{Thomas1997handbook}.  The state of the art, especially concerning the algebraic classification of automata, will appear in the forthcoming handbook~\cite{AutoMathAHandbook}.

\section{Automata and rational expressions}\label{section:automata}

\subsection{Operations on languages}

We describe here a collection of basic operations on languages, which will be building blocks in the characterization of the expressive power of automata.

Since languages over $A$ are subsets of $A^*$, we may of course consider the boolean operations: union, intersection and complement. 
The product operation on words can be naturally extended to languages: if $K$ and $L$ are languages over $A$,  we define their \emph{concatenation product}\index{concatenation product} $KL$ to be the set of all products of a word in $K$ followed by a word in $L$: 
$$KL = \{uv \mid u\in K\textrm{ and }v\in L\}.$$
We also use the power notation for languages: if $n> 0$, $L^n$ is the product $LL\cdots L$ of $n$ copies of $L$. We let $L^0 = \{\epsilon\}$. Note that if $n> 1$, $L^n$ differs from the set of $n$-th powers of the elements of $L$.
The \emph{iteration}\index{iteration} (or \emph{Kleene star}\index{star (Kleene)}) of a language $L$ is the language $L^* = \bigcup_{n\ge 0}L^n$.

Finally, we introduce a simple rewriting operation, based on the use of morphisms. If $A$ and $B$ are alphabets, a \emph{morphism}\index{morphism} from $A^*$ to $B^*$ is a mapping $\phi\colon A^* \rightarrow B^*$ such that
\begin{enumerate}
	\item  $\phi(\epsilon) = \epsilon$,

	\item  for all $u,v \in A^*$, $\phi(u v) = \phi(u) \phi(v)$.
\end{enumerate}
To specify such a morphism, it suffices to give the images of the 
letters of $A$. Then the image of a word $u\in A^{*}$, say $u=a_1 \cdots 
a_n$, is obtained by taking the concatenation of the images of the 
letters, $\phi(u)=\phi(a_1) \cdots \phi(a_n)$. That is, $\phi(a_1\cdots a_n)$ is obtained from $a_1\cdots a_n$ by substituting for each letter $a_i$ the word $\phi(a_i)$.
This operation naturally extends from words to 
languages: if $L\subseteq A^*$, then $\phi(L) = \{\phi(u) \mid 
u \in L\}$.

The consideration of these operations leads to the classical definition of \emph{rational}\index{rational!language}\index{language!rational} languages (also called \emph{regular}\index{regular!language}\index{language!regular} languages). The operations of union, concatenation and iteration are called the \emph{rational operations}\index{rational!operation}. A language over alphabet $A$ is called rational if it can be obtained from the letters of $A$ by applying (a finite number of) rational operations.

More formally, the class of rational languages over the alphabet $A$, denoted 
$\rat A^*$, is the least class of languages such that
\begin{enumerate}
 \item the languages $\emptyset$ and $\{a\}$ are rational for each letter 
 $a\in A$;
 \item if $K$ and $L$ are rational languages, then $K\cup L$, $KL$ 
 and $L^*$ are also rational.
\end{enumerate}

\begin{example}
The language $\Big(\big(a^{*}(ab)^{*}A^{*}\cap 
A^{*}(ba)^{*}\big)^{2}\Big)^{*}$ is rational. (Note that in order to lighten the notation, we write $a, b,$ etc., instead of $\{a\}$, $\{b\}.$)
  
The language $\{\epsilon\}$, containing just the empty word, is 
rational.  Indeed, it is equal to $\emptyset^*$.

Any finite language (that is, containing only finitely many 
words) is rational.

Let $a,b\in A$ be distinct letters.  It is instructive to show that the following languages are rational: (a) the set of all words  which do not contain two consecutive $a$; (b) the set of all words which contain the factor $ab$ but not the factor $ba$.
\end{example}

We also consider the \emph{extended rational operations}\index{rational!extended}: these are the rational operations, and the operations of intersection, complement and morphic image. A language is said to be \emph{extended rational} if it can be obtained from the letters of $A$ by applying (a finite number of) extended rational operations. The class of extended rational languages over $A$ is written $\extrat A^*$.

Of course, all rational languages are extended rational. The definition of extended rational languages offers more expressive possibilities but as we will see, they are not properly more expressive  than rational 
languages.

\subsection{Automata}

Let us start with a couple of examples.

\begin{example}\label{premier exemple}
A coffee machine delivers a cup of coffee for \euro{.25}.  It accepts 
only coins of \euro{.20}, \euro{.10} and \euro{.05}.  While determining whether it 
has received a sufficient sum, the machine is in one of six states, 
$q_0$, $q_{0.05}$, $q_{0.1}$, $q_{0.15}$, $q_{0.2}$ and $q_{0.25}$.  The names of 
the states correspond to the sum already received.  The machine 
changes state after a new coin is inserted, and the new state it 
assumes is a function of the value of the new coin inserted and of the 
sum already received. The latter information is encoded in the 
current state of the machine.

Here, the input word is the sequence of coins inserted, and the 
alphabet consists of three letters, {\tt w}, {\tt t} and {\tt f}, 
standing respectively for t\underline{\tt w}enty cents, \underline{\tt 
t}en cents and \underline{\tt f}ive cents.  The machine is 
represented in Figure~\ref{fig: coffee}.

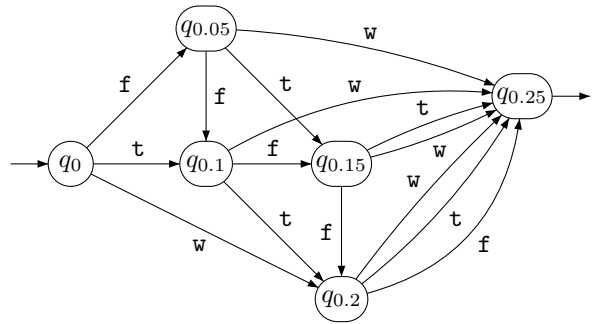
\begin{figure}[t]
\begin{center}
\begin{picture}(64,40)(-5,0)

\node[Nmarks=i,Nw=6.0,Nh=6.0](A)(0,18){$q_0$}
\node[Nw=8.0,Nh=6.0](B)(18,36){$q_{0.05}$}
\node[Nw=7.0,Nh=6.0](C)(18,18){$q_{0.1}$}
\node[Nw=8.0,Nh=6.0](D)(36,18){$q_{0.15}$}
\node[Nw=7.0,Nh=6.0](E)(36,0){$q_{0.2}$}
\node[Nmarks=f,Nw=8.0,Nh=6.0](F)(60,27){$q_{0.25}$}

\drawedge(A,B){{\tt f}}
\drawedge(A,C){\tt t}
\drawedge[ELside=r](A,E){\tt w}
\drawedge[curvedepth=2](B,F){\tt w}
\drawedge(B,D){\tt t}
\drawedge(B,C){\tt f}
\drawedge(C,D){\tt f}
\drawedge(C,E){\tt t}
\drawedge[curvedepth=4](C,F){\tt w}
\drawedge[curvedepth=1](D,F){\tt t}
\drawedge[ELside=r,curvedepth=-1](D,F){\tt w}
\drawedge[ELside=r](D,E){\tt f}
\drawedge[curvedepth=1](E,F){\tt w}
\drawedge[ELside=r,curvedepth=-2](E,F){\tt t}
\drawedge[ELside=r,curvedepth=-7](E,F){\tt f}

\end{picture}
\end{center}
\caption{The automaton of a (simplified) coffee machine}
\label{fig: coffee}
\end{figure}

The incoming arrow indicates the initial state of the machine 
($q_0$), and the outgoing arrow indicates the only accepting state
($q_{0.25}$), that is, the state in which the machine will indeed 
prepare a cup of coffee for you. Notice that the machine does not return change, but that it will accept 
sums up to \euro{.40}.
\end{example}

\begin{example}\label{deuxieme exemple}
Our second example (Figure~\ref{fig: mod 3}) reads an integer, given by its binary expansion and read from right to left, that is, starting with the bit of least weight. Upon reading this word on alphabet $\{0,1\}$, the automaton decides whether the given integer is divisible by 3 or not.

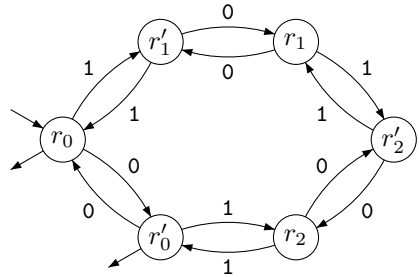
\begin{figure}[t]
\vspace*{3pt}
\begin{center}
\begin{picture}(55,31)(-5,0)

\node[Nmarks=if,iangle=150,fangle=210,Nw=6.0,Nh=6.0](A)(0,13){$r_0$}
\node[Nw=6.0,Nh=6.0](B)(13,26){$r'_1$}
\node[Nw=6.0,Nh=6.0](C)(31,26){$r_{1}$}
\node[Nw=6.0,Nh=6.0](D)(44,13){$r'_2$}
\node[Nw=6.0,Nh=6.0](E)(31,0){$r_{2}$}
\node[Nmarks=f,fangle= 210,Nw=6.0,Nh=6.0](F)(13,0){$r'_0$}

\drawedge[curvedepth=2](A,B){\small\tt1}
\drawedge[curvedepth=2](B,A){\small\tt1}

\drawedge[curvedepth=2](C,D){\small\tt1}
\drawedge[curvedepth=2](D,C){\small\tt1}

\drawedge[curvedepth=2](E,F){\small\tt1}
\drawedge[curvedepth=2](F,E){\small\tt1}

\drawedge[curvedepth=2](A,F){\small\tt0}
\drawedge[curvedepth=2](F,A){\small\tt0}

\drawedge[curvedepth=2](C,B){\small\tt0}
\drawedge[curvedepth=2](B,C){\small\tt0}

\drawedge[curvedepth=2](E,D){\small\tt0}
\drawedge[curvedepth=2](D,E){\small\tt0}

\end{picture}
\end{center}
\caption{An automaton to compute mod 3 remainders}
\label{fig: mod 3}
\end{figure}

For instance, consider the integer 19, in binary expansion {\tt 
10011}: our input word is {\tt 11001}.  It is read letter by 
letter, starting from the initial state (the state indicated by an 
incoming arrow, state $r_0$).  After each new letter is read, we 
follow the corresponding edge starting at the current state.  Thus, 
starting in state $r_0$, we visit successively the states $r'_1$, $r_0$, $r'_0$, $r_0$ again, and finally $r'_1$.  This state is not 
accepting (it is not marked with an outgoing edge), so the word {\tt 
11001} is not accepted by the automaton.  And indeed, 19 is not 
divisible by 3.

In contrast, 93 is divisible by 3, which is confirmed by running its binary expansion, namely {\tt 
1011101}, read from right to left, through the automaton: starting in state $r_0$, we end in state $r'_0$.

The reader will quickly see that this automaton is constructed in such a way that, if $n$ is an integer 
and $w_{n}$ is the binary expansion of 
$n$, then the state reached when reading $w_{n}$ from right to left, starting in state $r_{0}$, is $r_{k}$ (resp. $r'_k$) if $n$ is 
congruent to $k$ (mod 3) and $w_{n}$ has even (resp. odd) length.
\end{example}

We now turn to a formal definition. A (\emph{finite state}) \emph{automaton}\index{automaton} on alphabet 
$A$ is a 4-tuple $\calA = \automate$ where $Q$ is a finite set, 
called the set of \emph{states}\index{state}, $T$ is a subset of $Q\times A\times 
Q$, called the set of \emph{transitions}\index{transition}, and $I$ and $F$ are subsets 
of $Q$, called respectively the sets of \emph{initial states}\index{state!initial, final, accepting} and \emph{final states}. Final states are also called \emph{accepting 
states}.

For instance, the automaton of Example~\ref{premier exemple} uses a 
3-letter alphabet, $A = \{{\tt f},{\tt t},{\tt w}\}$.  Formally, 
it is the automaton $\calA = \automate$ given by $Q = \{q_0, q_{0.05}, 
q_{0.1}, q_{0.15}, q_{0.2}, q_{0.25}\}$, $I = \{q_0\}$, $F = \{q_{0.25}\}$ and $T$ is a 15-element subset of $Q \times A \times Q$ containing such triples as $(q_0,{\tt f},q_{0.05})$, $(q_{0.1},{\tt t},q_{0.2})$ or $(q_{0.2},{\tt w},q_{0.25})$.

As in our first examples, it is often convenient to represent an 
automaton $\calA = \automate$ by a labeled graph, 
whose vertices are the elements of $Q$ (the states) and whoses edges are 
of the form $q\buildrel a\over\longrightarrow q'$ if $(q,a,q')$ is a
transition, that is, if $(q,a,q')\in T$. The initial states are 
specified by an incoming arrow, and the final states are specified by 
an outgoing edge.

From now on, we will most often specify our automata by their 
graphical\break representations.

\begin{example}\label{dessin de A*abA*}
Here, the alphabet is $A = \{a,b\}$. Figure~\ref{fig: automate A*abA*} represents the automaton $\calA = \automate$ where $Q = \{1, 2, 3\}$, $I = \{1\}$, $F = \{3\}$ and
$$T = \{(1,a,1), (1,b,1), (1,a,2), (2,b,3), (3,a,3), (3,b,3)\}.$$
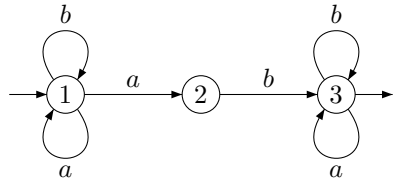
\begin{figure}[t]
\vspace*{3pt}
\begin{center}
\begin{picture}(46,16)(-5,-6)

\node[Nmarks=i,Nw=5.0,Nh=5.0](A)(0,0){1}
\node[Nw=5.0,Nh=5.0](B)(18,0){2}
\node[Nmarks=f,Nw=5.0,Nh=5.0](C)(36,0){3}

\drawloop[loopdiam=6](A){$b$}
\drawloop[loopdiam=6,loopangle=270](A){$a$}
\drawloop[loopdiam=6](C){$b$}
\drawloop[loopdiam=6,loopangle=270](C){$a$}
\drawedge(A,B){$a$}
\drawedge(B,C){$b$}

\end{picture}
\end{center}
\caption{An automaton accepting $A^*abA^*$}
\label{fig: automate A*abA*}
\end{figure}
\end{example}

\subsubsection{The language accepted by an automaton}

A \emph{path}\index{path} in automaton $\calA$ is a sequence of consecutive edges,
$$p = (q_{0},a_{1},q_{1})(q_{1},a_{2},q_{2})\ \cdots\ 
(q_{n-1},a_{n},q_{n}),$$
also drawn as
$$p = q_0\buildrel a_1\over\longrightarrow q_1\buildrel
a_2\over\longrightarrow q_2\ \ \cdots\ \buildrel
a_n\over\longrightarrow q_n.$$
Then we say that  $p$ is a path of \emph{length}\index{path!label, length} $n$ from $q_0$ to $q_n$, \emph{labeled} by the word $u = 
a_1a_2\cdots a_n$.  By convention, for each state $q$, 
there exists an \emph{empty path} from $q$ to $q$ labeled by the empty word.

For instance, in the automaton of Figure~\ref{fig: automate A*abA*}, the word $a^{3}ba$ labels exactly four paths: from 1 to 1, from 1 to 
2, from 1 to 3 and from 3 to 3.

A path $p$ is \emph{successful}\index{path!successful} if its initial state is in $I$ and its final state is in $F$. A word $w$ is \emph{accepted} (or \emph{recognized}) by $\calA$ if there exists a successful path in the automaton with label $w$.
And the \emph{language accepted}\index{language!accepted, recognized} (or \emph{recognized}) by $\calA$ is the set of labels of successful paths in $\calA$. It is denoted by $L(\calA)$. 
We say that $\calA$ \emph{accepts} (or \emph{recognizes}) $L(\calA)$.

For instance, the language of the automaton of Figure~\ref{fig: coffee} is finite, with exactly 27 words. The automaton of Figure~\ref{fig: automate A*abA*} accepts the
set of words in which at least one occurrence of $a$ is followed immediately by a $b$, namely $A^{*}abA^{*}$, where $A=\{a,b\}$.

Different automata may recognize the same language: if $\calA$ and $\calB$ are automata such that $L(\calA) = L(\calB)$, we say that $\calA$ and $\calB$ are \emph{equivalent}\index{automaton!equivalent}.

\begin{example}\label{A*abA* deterministe}
The language $A^{*}abA^{*}$, accepted by the automaton in Figure~\ref{fig: automate A*abA*}, is also recognized by the automaton in Figure~\ref{fig: automate deterministe A*abA*}
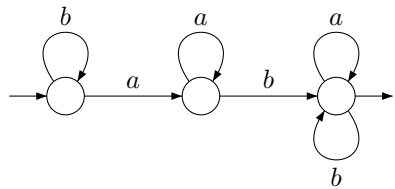
\begin{figure}[t]
\vspace*{3pt}
\begin{center}
\begin{picture}(46,16)(-5,-6)

\node[Nmarks=i,Nw=5.0,Nh=5.0](A)(0,0){}
\node[Nw=5.0,Nh=5.0](B)(18,0){}
\node[Nmarks=f,Nw=5.0,Nh=5.0](C)(36,0){}

\drawloop[loopdiam=6](A){$b$}
\drawloop[loopdiam=6](C){$a$}
\drawloop[loopdiam=6,loopangle=270](C){$b$}
\drawloop[loopdiam=6](B){$a$}
\drawedge(A,B){$a$}
\drawedge(B,C){$b$}

\end{picture}
\end{center}
\caption{Another automaton accepting $A^*abA^*$}
\label{fig: automate deterministe A*abA*}
\end{figure}
\end{example}

A language $L$  is said to be \emph{recognizable}\index{language!recognizable} if it is recognized by an automaton.

\subsubsection{Complete automata}

An automaton $\calA = \automate$ on alphabet $A$ is said to be
\emph{complete}\index{automaton!complete} if, for each state $q\in Q$ and each letter $a\in A$,
there exists at least one transition of the form $(q,a,q')$:  in graphical representation, this means that, for each letter of 
the alphabet, there is an edge labeled by that letter starting from 
each state. Naturally, this easily implies that, for each state 
$q$ and each word $w\in A^{*}$, there exists at least one path 
labeled $w$ starting at $q$. 

Every automaton can easily be turned into an equivalent complete automaton. If $\calA=\automate$ is not complete, the  \emph{completion}\index{completion} of $\calA$ is the automaton $\calA_{\rm comp} = 
(Q',T',I,F)$ given by $Q'= Q\cup\{z\}$, where $z$ is a new state not 
in $Q$, and $T'$ is obtained by adding to $T$ all triples 
$(z,a,z)$ ($a\in A$) and all triples $(q,a,z)$ ($q\in Q$, $a\in A$) 
such that there is no element of the form $(q,a,q')$ in $T$.

If $\calA$ is complete, we let $\calA_{\rm comp} = \calA$. It is immediate that, in every case, $\calA_{\rm comp}$ is complete and $L(\calA_{\rm comp})=L(\calA)$.

\begin{example}\label{exemple completion}
Let $A = \{a,b\}$. The automaton $\calB$ in Figure~\ref{fig: automata b*a*}, which accepts the language $b^*a^*$, is evidently not complete. The automaton $\calB_{\rm comp}$ is represented next to it.
\begin{figure}[t]
\vspace*{3pt}
\begin{center}
\begin{picture}(86,16)(-5,-6)

\put(-13,-8){$\calB$}
\node[Nmarks=if,fangle=270,Nw=5.0,Nh=5.0](A)(0,0){}
\node[Nmarks=f,fangle=270,Nw=5.0,Nh=5.0](B)(18,0){}

\drawloop[loopdiam=6](A){$b$}
\drawloop[loopdiam=6](B){$a$}
\drawedge(A,B){$a$}

\put(85,-8){$\calB_{\rm comp}$}
\node[Nmarks=if,fangle=270,Nw=5.0,Nh=5.0](AA)(40,0){}
\node[Nmarks=f,fangle=270,Nw=5.0,Nh=5.0](BB)(58,0){}
\node[Nw=5.0,Nh=5.0](CC)(76,0){$z$}

\drawloop[loopdiam=6](AA){$b$}
\drawloop[loopdiam=6](CC){$a$}
\drawloop[loopdiam=6,loopangle=270](CC){$b$}
\drawloop[loopdiam=6](BB){$a$}
\drawedge(AA,BB){$a$}
\drawedge(BB,CC){$b$}

\end{picture}
\end{center}
\caption{Two automata accepting $b^*a^*$}
\label{fig: automata b*a*}
\end{figure}
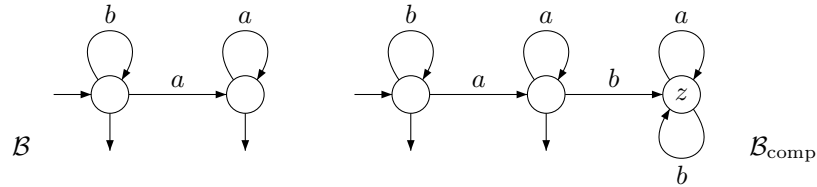
\end{example}





\subsubsection{Trim automata}\label{sec: trim}

A complete automaton reads its entire input before deciding to accept or reject it: whatever input it receives, there is a transition that can be followed. However, we have seen that in the completion $\calA_{\rm comp}$ of a non-complete 
automaton $\calA$, state $z$ does not participate in any successful 
path: it is in a way a useless state. {\it Trimming} an automaton removes such useless states; it is, in a sense, the opposite of completing an automaton, and aims at producing a more concise device.

A state $q$ of an automaton $\calA$ is said to be
\emph{accessible}\index{state!accessible} if there exists a path in $\calA$ starting from some initial state and ending at $q$. State $q$ is \emph{co-accessible} if there exists a path in $\calA$ starting from $q$ and ending at some final state. Observe that a state is both accessible and co-accessible if and only if it is visited by at least one successful path.

The automaton $\calA$ itself is \emph{trim}\index{automaton!trim} if all its states are both
accessible and co-accessible: in a trim automaton, each state is useful, in the sense that it is used in accepting some word of the language $L(\calA)$.

Of course, every automaton $\calA$ is equivalent to a trim one, written $\calAem$, obtained by restricting $\calA$ to its accessible and co-accessible states and to the transitions between them.


Interestingly, $\calAem$ can be constructed efficiently, using breadth-first search. One first computes the accessible states of $\calA$, by letting $Q_0 = I$ (the initial states are 
certainly accessible) and by computing iteratively
$$Q_{n+1} \enspace=\enspace Q_n \cup \bigcup_{q\in Q_n, a\in A} \{q'\in Q \mid (q,a,q')\in T\}.$$
%
%
One verifies that the elements of $Q_n$ are the states that can be reached from an initial state, reading a word of length at most $n$; and that if two consecutive sets $Q_n$ and $Q_{n+1}$ are equal, then $Q_n = Q_m$ for all $m\ge n$, and $Q_n$ is the set of accessible states of $\calA$. In particular, the set of accessible states is computed in at most $|Q|$ steps.

A similar procedure, starting from the final states instead of the initial states, and working in reverse, produces in at most $|Q|$ steps the set of co-accessible states of $\calA$. The automaton $\calAem$ is then immediately constructed.

\begin{remark}\label{addl rk 1}
The construction of $\calAem$, or indeed, just of the set of accessible states of $\calA$ provides an efficient solution of the \emph{emptiness problem}\index{emptiness problem}: given an automaton $\calA$, is the language $L(\calA)$ empty? that is, does $\calA$ accept at least one word?

Indeed, $\calA$ recognizes the empty set if and only if no final state is accessible: in order to decide the emptiness problem for automaton $\calA$, it suffices to construct the set of accessible states of $\calA$ and verify whether it contains a final state. This yields an $\O(|Q|^2|A|)$ algorithm.
\end{remark}

\subsubsection{Epsilon-automata}\label{sec: epsilon}

It is sometimes convenient to extend the notion of automata to the so-called \emph{$\epsilon$-automata}\index{automaton!$\epsilon$-}: the difference from ordinary automata is that we also allow $\epsilon$-labeled transitions, of the form $(p,\epsilon,q)$ with $p,q\in Q$.

\begin{proposition}\label{equivalence epsilon}
Every $\epsilon$-automaton is equivalent to an ordinary automaton.
\end{proposition}

\sketch
Let $\calA =  \auto$ be an $\epsilon$-automaton, and let $\calR$ be the relation on $Q$ given by $p \calR q$ if there exists a path from $p$ to $q$ consisting only of $\epsilon$-labeled transitions (that is: $\calR$ is the reflexive transitive closure of the relation defined by the $\epsilon$-labeled transitions of $\calA$).

Let $\calA'$ be the (ordinary) automaton given by the tuple $(Q,T',I',F)$ with
\begin{align*}
T' &= \big\{(p,a,q) \mid (p,a,q') \in T\textrm{ and }q' \calR q\textrm{ for some $q'\in Q$}\big\} \\
I' &= \big\{q \mid p\calR q \textrm{ for some $p\in I$}\big\}.
\end{align*}
Then $\calA'$ is equivalent to $\calA$.
\eosketch

\subsection{Deterministic automata}

\begin{example}\label{det vs ndet, A^*abA^*}
Consider the automaton of Figure~\ref{fig: automate A*abA*}, say $\calA$, and the automaton  $\calB$ of Figure~\ref{fig: automate deterministe A*abA*}. Both recognize the  language, $L = A^{*}abA^{*}$, but there is an important, qualitative difference beween them.



We have defined automata as \emph{nondeterministic} computing devices:  given a state and an input letter, there may be several possible choices for the next state.  Thus an input word might be associated with many different computation paths, and the word is accepted if one of these paths ends at an accepting state.  In contrast, $\mathcal{B}$ has the convenient property that each input word labels at most one computation path.
\end{example}

These remarks are formalized in the following definition. An automaton $\calA = \automate$ is said to be \emph{deterministic}\index{automaton!deterministic} if it has exactly one initial state, and if, for each letter $a$ and  for all states $q, q', q''$,
$$(q,a,q'),\ (q,a,q'')\in T\qquad\Longrightarrow\qquad q'=q''.$$
Thus, of the  automata in Figures~\ref{fig: automate A*abA*} and~\ref{fig: automate deterministe A*abA*}, the second one is deterministic, and the first is non-deterministic.

This definition imposes a certain condition of uniqueness on transitions, that is, on paths of length 1. This property is then extended to longer paths by a simple induction.

\begin{proposition}\label{unicite chemin}
Let $\calA$ be a deterministic automaton and let $w$ be a word.
\begin{description}
\item[\rm(1)] For each state $q$ of $\calA$, there exists at most one 
path labeled $w$ starting at $q$.
\item[\rm(2)] If $w\in L(\calA)$, then $w$ labels exactly one 
successful path.
\end{description}
\end{proposition}

In particular, we can represent the set of transitions of a deterministic automaton $\calA =\automate$ by a \emph{transition function}\index{transition!function}: the (possibly partial) function $\delta\colon Q\times A \rightarrow Q$ which maps each pair $(q,a)\in Q\times A$ to the state $q'$ such that $(q,a,q')\in T$ (if it exists). This function is then naturally extended to the set $Q\times A^{*}$: if $q\in Q$ and $w\in A^*$, $\delta(q,w)$ is the state $q'$ such that  there exists a path from $q$ to $q'$ labeled by $w$ in $\calA$ (if  such a state exists). In the sequel, deterministic automata will be specified as 4-tuples $(Q,\delta,i,F)$ instead of the corresponding $(Q,T,\{i\},F)$. We note the following elementary characterization of $\delta$.

\begin{proposition}\label{prop: transition function}
Let $\calA = (Q,\delta,i,F)$ be a deterministic automaton.  Then we have
\begin{align*}
\delta(q, \epsilon) & = q;\\
\delta(q,ua) & = \begin{cases}
				\delta(\delta(q, u), a) & \textrm{if both $\delta(q, u)$ and $\delta(\delta(q, u), a)$ exist,}\cr
				\textrm{undefined} & \textrm{otherwise;}
			\end{cases} \\
u \in L(\calA) & \textrm{ if and only if $\delta(i,u) \in F$.}                             
\end{align*}
for each state $q$, each word $u\in A^*$ and each letter $a\in A$.
\end{proposition}

Again, it turns out that every automaton is equivalent to a deterministic automaton.  This deterministic automaton can be effectively constructed, although the algorithm -- the so-called \emph{subset construction}\index{subset!construction} -- is more complicated than those used to construct complete or trim automata.

Let $\calA = \automate$ be an automaton. The \emph{subset transition function}\index{subset!transition function} of $\calA$ is the function $\delta\colon
{\cal P}(Q)\times A \rightarrow {\cal P}(Q)$ defined, for each
$P\subseteq Q$ and each $a\in A$ by
$$\delta(P,a) = \{q\in Q \mid \exists p\in P,\ (p,a,q)\in T\}.$$
Thus, $\delta(P,a)$ is the set of states of $\calA$ which can be 
reached by an $a$-labeled transition, starting from an element of $P$.
The \emph{subset automaton}\index{subset!automaton}\index{automaton!subset} of $\calA$ is $\calA_{\rm sub} = (\calP(Q), \delta,I,F_{\rm sub})$ where $F_{\rm sub} = \{P\subseteq Q   \mid P\cap F\ne\emptyset\}$.

The automaton $\calA_{\rm sub}$ is deterministic and complete by construction, and the subset transition function of $\calA$ is the transition function of $\calA_{\rm sub}$. Moreover, if $\calA$ has $n$ states, then $\calA_{\rm sub}$ has $2^n$ states.

\begin{example} \label{ex automate parties}
The subset automaton of the non-deterministic automaton of Figure~\ref{fig: automate A*abA*} is given in Figure~\ref{fig: subset automaton}. Notice that the states of the second row are not accessible.
\begin{figure}[t]
\vspace*{3pt}
\begin{center}
\begin{picture}(64,37)(-5,-9)

\node[Nmarks=i,Nw=7.0,Nh=7.0](A)(0,18){$\{1\}$}
\node[Nw=10.0,Nh=7](B)(20,18){$\{1,2\}$}
\node[Nmarks=f,fangle=45,Nw=10.0,Nh=7](C)(40,18){$\{1,3\}$}
\node[Nmarks=f,fangle=45,Nw=13.0,Nh=7](D)(60,18){$\{1,2,3\}$}
\node[Nw=7.0,Nh=7.0](E)(0,0){$\emptyset$}
\node[Nw=7.0,Nh=7.0](F)(20,0){$\{2\}$}
\node[Nmarks=f,fangle=-45,Nw=7.0,Nh=7.0](G)(40,0){$\{3\}$}
\node[Nmarks=f,fangle=-45,Nw=10.0,Nh=7](H)(60,0){$\{2,3\}$}

\drawloop[loopdiam=6](A){$b$}
\drawedge(A,B){$a$}
\drawloop[loopdiam=6](B){$a$}
\drawedge(B,C){$b$}
\drawedge[curvedepth=3](C,D){$a$}
\drawloop[loopdiam=6](C){$b$}
\drawedge[curvedepth=3](D,C){$b$}
\drawloop[loopdiam=6](D){$a$}

\drawloop[loopdiam=6](E){$a$}
\drawloop[loopdiam=6,loopangle=270](E){$b$}
\drawedge[ELside=r](F,E){$a$}
\drawedge(F,G){$b$}
\drawloop[loopdiam=6](G){$a$}
\drawloop[loopdiam=6,loopangle=270](G){$b$}
\drawedge[ELside=r,curvedepth=-3](H,G){$b$}
\drawedge[curvedepth=3](H,G){$a$}

\end{picture}
\end{center}
\caption{The subset automaton of the automaton in Figure~\ref{fig: automate A*abA*}}
\label{fig: subset automaton}
\end{figure}
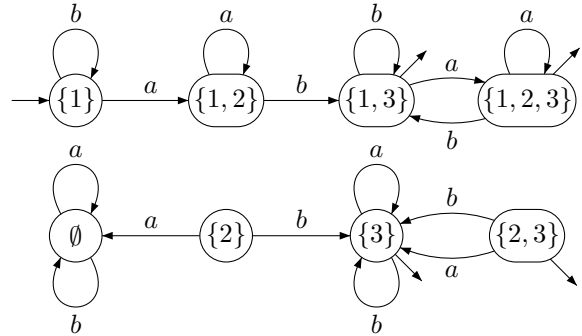
\end{example}

\begin{proposition}\label{prop parties}
The automata $\calA$ and $\calA_{\rm sub}$ are equivalent.
\end{proposition}

\sketch
Let $\calA = \automate$. One shows by induction on $|w|$ that for all $P \subseteq Q$ and $w\in A^*$, $\delta(P,w)$ is the set 
of all states $q\in Q$ such that $w$ labels a path in $\calA$ 
starting at some state in $P$ and ending at $q$.

Therefore, a word $w$ is accepted by $\calA$ if and only if at least 
one final state lies in the set $\delta(I,w)$, if and only if 
$\delta(I,w)\in F_{\rm sub}$, if and only if $w$ is accepted 
by  $\calA_{\rm sub}$. This concludes the proof.  
\eosketch

In general, the subset automaton is not trim (see 
Example~\ref{ex automate parties}) and we can find a deterministic 
automaton smaller than $\calA_{\rm sub}$, which still recognizes the 
same language as $\calA$, namely by trimming $\calA_{\rm sub}$. Observe that in the proof of Proposition~\ref{prop parties}, the only useful states of $\calA_{\rm sub}$ are those of the form $\delta(I,w)$, that 
is, the accessible states of $\calA_{\rm sub}$.

We define the \emph{determinized automaton}\index{automaton!determinized}\index{determinization} of $\calA$ to be  $\calA_{\rm det} = \big(\calA_{\rm  sub})_{\rm trim}$. This automaton is equivalent to $\calA$.

\begin{example} \label{ex determinise}
The determinized automaton of the non-deterministic automaton of Figure~\ref{fig: automate A*abA*} consists of the first row of states in Figure~\ref{fig: subset automaton} (see Example~\ref{ex automate parties}).
\end{example}

An obstacle in the computation of $\calA_{\rm det}$ is the explosion in the number of states: if $\calA$ has $n$ states, then $\calA_{\rm sub}$ has $2^n$ states. The determinized automaton $\calA_{\rm det}$ may well have exponentially many states as well, but it sometimes has fewer. Therefore, it makes sense to try and compute $\calA_{\rm det}$ directly, in time proportional to its actual number of states, rather than first constructing the exponentially large automaton $\calA_{\rm sub}$ and then trimming it.

This can be done using the same ideas as in the construction of $\calA_{\rm trim}$ in Section~\ref{sec: trim}. One first constructs $\calB$, the accessible part of $\calA_{\rm sub}$, starting with the initial state of $\calA_{\rm sub}$, namely $I$. Then for each constructed state $P$ and each letter $a$, we construct $\delta(P,a)$ and the transition $(P,a,\delta(P,a))$. And we stop when no new state arises this 
way.

The second step consists in finding the co-accessible part of $\calB$, using the method in Section~\ref{sec: trim}.

\begin{example}\label{determ exponentielle 1}
Let $A = \{a,b\}$, let $n\ge 2$, and let $L = A^*aA^{n-2}$. Then $L$ is accepted by a  non-deterministic automaton $\calA$ with $n$ states. However, any deterministic automaton accepting $L$ must have at least $2^{n-1}$ states.  To see this, suppose that $(Q,\delta,i,F)$ is such a deterministic automaton. Let $u, v$ be distinct words of length $n-1$.  Then one of the words (let us say $u$) contains an $a$ in a position in which $v$ contains the letter $b$.  Thus $u=u'ax$, $v=v'by$, where $|x|=|y|$.  Let $w$ be any word of length $n-2-|x|$.  Then $uw\in L$, $vw\notin L$.  It follows that $\delta(i,u)\neq \delta(i,v)$ and thus there are at least as many states as there are words of length $n-1$.  This shows that the exponential blowup in the number of states in the subset construction cannot in general be reduced.
%
\end{example}

\section{Logic: B\"uchi's sequential calculus}\label{section:logic}

Let us start with an example.

\begin{example}\label{first logic example}
Recall that $\et$ is the logical conjunction, which reads ``AND''. And $\ou$ is the logical disjunction, which reads ``OR''. We will consider formulas such as 
$$
\exists x \exists y\ (x < y) \et R_ax \et R_by.
$$
This formula has the following interpretation on a word $u$: there exist two natural numbers $x < y$ such that, in $u$, the letter in position $x$ is an $a$ and the letter in position $y$ is a $b$. Thus this formula specifies a language: the set of all words $u$ in which this formula holds, namely $A^*aA^*bA^*$.
\end{example}

\subsection{First-order formulas}\label{sec: FO formulas}

Let us now formalize this point of view on languages.

\subsubsection{Syntax}

The formulas of B\"uchi's sequential calculus\index{B\"uchi's sequential calculus} use the usual logical symbols ($\land$, $\lor$, $\neg$ for the negation), the equality symbol $=$, the constant symbol \vrai,  the quantifiers $\exists$ and $\forall$, variable symbols ($x, y, z, \ldots$) and parentheses. They also use specific, non-logical symbols: binary relation symbols $<$ and $S$, and unary relation symbols $R_{a}$ (one for each letter $a\in A$).

For convenience, we may assume that the variables are drawn from a fixed, countable, set of variables.

The \emph{atomic formulas}\index{formula!atomic} are the formulas of the form \vrai, $x = y$, $x < y$, $S(x,y)$, and $R_ax$, where $x$ and $y$ are variables and $a\in A$.

The \emph{first-order formulas}\index{formula!first-order} are defined as follows:
\begin{itemize}
\item Atomic formulas are first-order formulas,

\item If $\varphi$ and $\psi$ are first-order formulas, then $(\neg\varphi)$, $(\phi \et \psi)$ and $(\phi \ou \psi)$ are first-order formulas,

\item If $\varphi$ is a first-order formula and if $x$ is a variable, then
$(\exists x\ \varphi)$ and $(\forall x\ \varphi)$ are first-order formulas.
\end{itemize}

\begin{remark}
As is usual in logic, we will limit the usage of parentheses in our notation of formulas, to what is necessary for their proper parsing, writing for instance $\forall x\ R_ax$ instead of $(\forall x\ (R_ax))$.
\end{remark}

Certain variables appear after a quantifier (existential or universal): occurrences of these variables within the scope of the quantifier are said to be \emph{bound}\index{variable!free, bound}. Other occurrences are said to be \emph{free}. A precise, recursive, definition of the set $FV(\varphi)$ of the free variables of a formula $\phi$ is as follows:
\begin{itemize}
  \item If $\varphi$ is atomic, then $FV(\varphi)$ is the set of all variables occurring in $\varphi$,
  
  \item $FV(\neg\varphi)=FV(\varphi)$,
  
  \item $FV(\varphi\et\psi) = FV(\varphi\ou\psi)=FV(\varphi)\cup FV(\psi)$,
  
  \item $FV(\exists x\ \varphi)= FV(\forall x\ \varphi)=FV(\varphi)\setminus \{x\}$.
\end{itemize}
A formula without free variables is called a \emph{sentence}\index{sentence}.

\subsubsection{Interpretation of formulas}\label{sec: eval structures}\label{sec: eval words}

In B\"uchi's sequential calculus, formulas are interpreted in words: each word $u$ of length $n \ge 0$ determines a \emph{structure}\index{structure} (which we abusively denote by $u$) with \emph{domain}\index{domain} $\Dom(u) = \{0, \ldots ,n-1\}$ ($\Dom(u) = \emptyset$ if $u = \epsilon$). $\Dom(u)$ is viewed as the set of positions in the word $u$ (numbered from 0).

The symbol $<$ is interpreted in $\Dom(u)$ as the usual order (as in $(2<4)$ and $\neg(3<2)$). The symbol $S$ is interpreted as the \emph{successor}\index{successor} symbol: if $x,y\in\Dom(u)$, then $S(x,y)$ if and only if $y=x+1$. Finally, for each letter $a\in A$, the unary relation symbol $R_a$ is interpreted as the set of positions in $u$ that carry an $a$ (a subset of $\Dom(u)$).

\begin{example}
If $u = abbaab$, then $\Dom(u) = \{0, 1, \ldots ,5\}$, $R_a =\{0,3,4\}$ and  $R_b = \{1,2,5\}$.
\end{example}

A \emph{valuation}\index{valuation} on $u$  is a mapping $\nu$ from a set of variables into the domain $\Dom(u)$. It will be useful to have a notation for small modifications of a valuation: if $\nu$ is a valuation and $d$ is an element of $\Dom(u)$, we let $\nu[x\mapsto d]$ be the 
valuation $\nu'$ defined by extending the domain of $\nu$ to include the variable $x$ and setting
$$\nu'(y)=\begin{cases}\nu(y) & \textrm{ if $y\not= x$,} \\
               d      &\textrm{ if $y=x$.} \end{cases}$$
If $\varphi$ is a formula, $u\in A^*$ and $\nu$ is a valuation on $u$ whose domain includes the free variables of $\varphi$, then we define $u,\nu \models\varphi$ (and say that the valuation $\nu$ \emph{satisfies}\index{satisfaction} $\varphi$ in 
$u$, or equivalently $u,\nu$ satisfies $\varphi$) as follows:
\begin{itemize}
  \item $u,\nu \models (x=y)$ (resp. $(x<y)$, $S(x,y)$, $R_ax$) if and only if $\nu(x)=\nu(y)$ (resp. $\nu(x)<\nu(y)$, $S(\nu(x),\nu(y))$, $R_a\nu(x)$) in $\Dom(u)$;
  
  \item $u,\nu \models\neg\varphi$ if and only if it is not true that 
$u,\nu \models\varphi$;
  
  \item $u,\nu \models (\varphi\ou \psi)$ (resp. $(\varphi\et \psi)$) if and only if at least one (resp. both) of $u,\nu \models 
\varphi$ and $u,\nu \models \psi$ holds (resp. hold);
  
  \item $u,\nu \models(\exists x\,\varphi)$ if and only if there exists $d\in \Dom(u)$ such that $u,\nu[x\mapsto d] \models \varphi$;
  
  \item $u,\nu \models(\forall x\,\varphi)$ if and only if, for each $d\in \Dom(u)$, $u,\nu[x\mapsto d] \models \varphi$.
\end{itemize}

Note that the truth value of $u,\nu \models\varphi$ depends only on the values assigned by $\nu$ to the free variables of 
$\varphi$. In particular, if $\phi$ is a sentence, then there is a valuation $\mu$ with an empty domain. We say that \emph{$\varphi$ is satisfied by $u$} (or  \emph{$u$ satisfies $\varphi$}), and we write $u \models \varphi$ for  $u,\mu \models\varphi$. Thus each sentence $\phi$ defines a language: the set $L(\phi)$ of all words such that $u \models \phi$.  Note that this interpretation makes sense even if $u$ is the empty word, for then the valuation $\mu$ is still defined:  Every sentence beginning with a universal quantifier is satisfied by $\epsilon$, and no sentence beginning with an existential quantifier is satisfied by $\epsilon$. An early example was given in Example~\ref{first logic example}, 

\begin{remark}
Two sentences $\varphi$ and $\psi$ are said to be \emph{logically equivalent} if they are satisfied by the same structures. We will use freely the classical logical equivalence results, such as the logical equivalence of $\varphi \et \psi$ and $\neg(\neg\varphi\ou\neg\psi)$, or the logical equivalence of $\forall x\ \varphi$ and  $\neg(\exists x\ \neg\varphi)$. We will also use the implication and bi-implication notation: $\phi \rightarrow \psi$ stands for $\neg \phi \ou \psi$ and $\phi \leftrightarrow \psi$ stands for $(\phi \to \psi) \et (\psi \to \phi)$.
%

%
\end{remark}

\begin{example}\label{ex logical aA^*}
Let $\phi$ and $\psi$ be the following formulas.
\begin{align*}
\phi &= \exists x\ \Big(\big(\forall y\ \neg(y < x)\big) \land R_ax\Big) \\
\psi &= \forall x\ \Big(\big(\forall y\ \neg(y < x)\big) \rightarrow R_ax\Big).
\end{align*}
The sentence $\phi$ states that there exists a position with no strict predecessor, containing an $a$, while $\psi$ states that every such position contains an $a$.  The latter sentence, like all universally quantified first-order sentences, is vacuously satisfied by the empty string. Thus $L(\phi) = aA^*$ and $L(\psi) = aA^* \cup \{\epsilon\}$.
\end{example}

The \emph{first-order logic}\index{logic!first-order} of the  \emph{linear order} (resp. of the \emph{successor}), written $\FO(<)$ (resp. $\FO(S)$) is the fragment of the first-order logic described so far, where formulas do not use the symbol $S$ (resp. $<$).

\subsection{Monadic second-order formulas}

In \emph{monadic second-order logic}\index{logic!monadic second-order}, we add a new type of variable to first-order logic, called \emph{set variables}\index{variable!set} and usually denoted by upper case letters, e.g. $X,Y,$ \dots\ The atomic formulas of monadic second-order are the atomic formulas of first-order logic, and the formulas of the form $(Xy)$, where $X$ is a set variable and $y$ is an ordinary variable.

The recursive definition of \emph{monadic second-order formulas}\index{formula!monadic second-order}, starting from the atomic formulas, closely resembles that of first-order formulas: it uses the same rules given in Section~\ref{sec: FO formulas}, and the additional rule:
\begin{itemize}
\item If $\varphi$ is a monadic second-order formula and $X$ is a set variable, then $(\exists X\varphi)$ and $(\forall X\varphi)$ are monadic second-order formulas.
\end{itemize}
The notion of free variables is extended in the same fashion.

The interpretation of monadic second-order formulas also requires an extension of the definition of a valuation on a word $u$: a \emph{monadic second-order valuation}\index{valuation!monadic second-order} is a mapping $\nu$ which associates with each first-order variable an element of the domain $\Dom(u)$, and with each set variable, a subset of $\Dom(u)$.

If $\nu$ is a valuation, $X$ is a set variable, and $R$ is a subset of 
$\Dom(u)$, we denote by $\nu[X\mapsto R]$ the valuation obtained from $\nu$ by mapping $X$ to $R$ (see Section~\ref{sec: eval structures}).

With these definitions, we can recursively give a meaning to the notion that a valuation $\nu$ satisfies a formula $\phi$ in a word $u$ ($u,\nu \models \phi$): we use again the rules given in Section~\ref{sec: eval structures}, to which we add the following:
\begin{itemize}
  \item $u,\nu \models (Xy)$ if and only if $\nu(y) \in \nu(X)$;
  
  \item $u,\nu \models(\exists X\varphi)$ (resp. $(\forall X\varphi)$) if and only if there exists $R\subseteq \Dom(u)$ such that (resp. for each $R \subseteq \Dom(u)$) $u,\nu[X\mapsto R] \models \varphi$.
\end{itemize}
Note that the empty set is a valid assignment for a set variable: the empty word may satisfy monadic second order variables even if they start with an existential set quantifier.

B\"uchi's sequential calculus (see Section~\ref{sec: eval words}) is thus extended to include monadic second-order formulas. We denote by $\MSO(<)$ (resp. $\MSO(S)$) the fragment of monadic second-order logic, where formulas do not use the symbol $S$ (resp. $<$). Of course, $\FO(<)$ and $\FO(S)$ are subsets of $\MSO(<)$ and $\MSO(S)$, respectively.

\begin{example}
Inspecting the following $\MSO(<)$ sentence,
\begin{align*}
\phi = \exists X\ \quad \bigl[\forall x\ &(Xx\leftrightarrow ((\forall y\ \neg(x<y)) \ou (\forall y\ \neg(y<x))))\\
\et\ & \forall x\ (Xx\rightarrow R_{a}x) \
\et\ \exists x\ Xx\bigr].
\end{align*}
one can see that the elements of $X$ must be the first and last positions of the word in which we interpret $\phi$, so $L(\phi)= aA^*\cap A^*a$. This language can also be described by a first order sentence, see Example~\ref{ex logical aA^*}, that is: this formula is equivalent to a first-order formula.
\end{example}

\begin{example}
We now consider the more complex formula
\begin{align*}
\phi = \exists X\ \quad &
\bigl((\forall x\ \forall y\ ((x<y)\et(\forall z\ \neg((x<z)\et(z<y))))\rightarrow 
(Xx\leftrightarrow\neg Xy))\\
\et\ & (\forall x\ (\forall y\ \neg(y<x))\rightarrow Xx)\\
\et\ & (\forall x\ (\forall y\ \neg(x<y))\rightarrow \neg Xx)\bigr).
\end{align*}
The formula $\phi$ states that there exists a set $X$ of positions in the 
word, such that a position is in $X$ if and only if the next position is not in $X$ (so $X$ has every other position), and the first position is in $X$, and the last position is not in $X$. Thus $L(\phi)$ is the set of words of
even length. It is an easy consequence of the results of Section~\ref{sec: FO SF Ap} that this language cannot be described by a first-order formula.
\end{example}

The successor relation can be expressed in $\FO(<)$: $S(x,y)$ is logically equivalent to the following formula:
$$\quad(x<y)\ \et\ \forall z\ ((x<z) \rightarrow ((y=z) \ou (y<z))).$$
In a weak converse, the order relation $<$ can be expressed in $\MSO(S)$: the formula $x < y$ is equivalent to:
$$\exists X\ \bigl(Xy \et \neg Xx\ \et [\forall z\ \forall t\ ((Xz \et S(z,t))\rightarrow Xt)]\bigr).$$
It follows that $\MSO(<)$ and $\MSO(S)$ have the same expressive power.

\begin{proposition}
A language can be defined by a sentence in $\MSO(S)$, if and only if it can be defined by a sentence in $\MSO(<)$.
\end{proposition}

However, the order relation $<$ cannot be expressed in $\FO(S)$. This is a non-trivial result; for a proof, see \cite{Straubing:1994}.

\begin{proposition}
If a language can be defined by a sentence in $\FO(S)$, then it can be defined by a sentence in $\FO(<)$. The converse does not hold.
\end{proposition}

\section{The Kleene-B\"uchi theorem}\label{section:kleene-buchi}

In this section, we prove the following theorem, a combination of the classical Kleene and B\"uchi theorems.

\begin{theorem}
  Let $L$ be a language in $A^*$. The following conditions are equivalent:
  \begin{enumerate}
	\item $L$ is defined by a sentence in $\MSO(<)$;
	\item $L$ is accepted by an automaton;
	\item $L$ is extended rational;
	\item $L$ is rational.
  \end{enumerate}
\end{theorem}

\subsection{From automata to monadic second-order formulas}

Let ${\cal A}=(Q,i,\delta,F)$ be a deterministic automaton. The idea is to associate with each state $q\in Q$ a second order variable $X_q$, to encode the set of positions in which a given path visits state $q$. What we need to express about the sets $X_{q}$ is the following:
\begin{itemize}
  \item the sets $X_q$ form a partition of the set of all positions (at each point in time, the automaton must be in one and exactly one state);
  
  \item if a path visits state $q$ at time $x$, state $q'$ at time $x+1$ and if the letter in position $x+1$ is an $a$, then $\delta(q,a)=q'$;
\end{itemize}
This analysis leads to the following formula. For convenience, let $Q$ be the set $\{q_{0},q_{1},\ldots,q_{n}\}$, with initial state $i=q_{0}$. We also use the shorthand $\min$ and $\max$ to designate the first and last positions: this is acceptable as these positions can be expressed by $\FO(S)$-formulas. For instance, $R_a\min$ stands for $\forall x\,(\forall y\ \neg S(y,x) \rightarrow R_a x)$; and 
$X\max$ stands for $\forall x\,(\forall y\ \neg S(x,y) \rightarrow X x)$.
\begin{align*}
\exists X_{q_{0}}&\enspace \exists X_{q_{1}}\enspace \cdots \enspace \exists X_{q_{n}} \\
&\Biggl( \bigwedge_{q\not= q'} \neg\exists x\ (X_qx\et X_{q'}x) \qquad\et \qquad \forall x\ \bigvee_{q} X_qx \\
\et \quad & \forall x\ \forall y\ \Big[S(x,y) \rightarrow \bigvee_{q\in Q,\ a\in A} \bigl(X_qx\et R_ay\et X_{\delta(q,a)}y\bigr)\Big] \\
\et \quad & \bigwedge_{a\in A} \bigl(R_a\min \rightarrow X_{\delta(q_0,a)}\min\bigr)\  \et\ \Bigl(\bigvee_{q\in F} X_q\max\Bigr)\Biggr).
\end{align*}
This sentence is actually verified by the empty word, so the language it defines coincides with  $L(\calA)$ on $A^+$. If $q_0\in F$, it accurately defines $L(\calA)$. But if $q_0\not\in F$, we must consider the conjunction of this sentence with $\exists x\ \vrai$.

This is a sentence in $\MSO(S,<)$ but as we know, it is logically equivalent to one in $\MSO(<)$. Note that it is in fact an existential monadic second order sentence, that is, the second-order quantifications are all existential.

\subsection{From formulas to extended rational expressions}\label{sec: formulas 2 expressions}

The proof that an $\MSO(<)$-definable language can be described by an extended rational expression, is more complex. The reasoning is by 
induction on the recursive definition of formulas. Instead of associating a language only with sentences (formulas without free variables), we will associate languages with all formulas but these languages will be over larger alphabets, which allow us to encode valuations.

\subsubsection{The auxiliary alphabets $B_{p,q}$}\label{Bpq}

Let $p, q \ge 0$ and  let $B_{p,q}=A\times\{0,1\}^p\times\{0,1\}^q$. A word over the alphabet $B_{p,q}$ can be identified with a sequence $(u_0,u_1,\ldots,u_p,u_{p+1}, \ldots,u_{p+q})$ where $u_0\in A^*$, $u_1,\ldots,u_p,u_{p+1}, \ldots,u_{p+q}\in \{0,1\}^*$ and all the $u_{i}$ have the same length.

Let $K_{p,q}$ consist of the empty word and the words in $B_{p,q}^+$ such that each of the components $u_1,\ldots,u_p$  contains exactly one occurrence of $1$.  Thus each of these components really designates \emph{one} position in the word $u_{0}$, and each of the components $u_{p+1},\ldots, u_{p+q}$ designates a set of positions in $u_0$.

\begin{example}\label{ex: B32}
If $A=\{a,b\}$, the following  is a word in $K_{2,1}$:
$$\begin{matrix}
u_{0} \qquad & a & b & a & a & b & a & b \cr
\relax & \cr
u_{1} \qquad & 0 & 0 & 0 & 0 & 1 & 0 & 0 \cr
u_{2} \qquad & 0 & 0 & 1 & 0 & 0 & 0 & 0 \cr
\relax & \cr
u_{3} \qquad & 0 & 1 & 1 & 0 & 0 & 1 & 1 \cr
\end{matrix}$$
Its components $u_1$ and $u_2$ designate positions 4 and 2, respectively, and its component $u_3$ designates the set $\{1,2,5,6\}$.
\end{example}

The languages $K_{p,q}$ are extended rational.  Indeed, for $1\leq i\leq p$, let $C_i$ be the set of elements $(b_0,b_1,\ldots ,b_{p+q}) \in B_{p,q}$ such that $b_i=1$.  Then $K_{p,q}$ is the set of words in $B_{p,q}^*$ which contain at most one letter in each $C_i$:
$$K_{p,q}= \bigl\{\epsilon\bigr\}\enspace \cup\enspace \bigcap_{1\leq i\leq p} (B_{p,q} \setminus C_i)^*C_i(B_{p,q} \setminus C_i)^* = B_{p,q}^* \setminus\bigcup_{1\leq i\leq p} B_{p,q}^*C_iB_{p,q}^*C_iB_{p,q}^*.$$
%

\subsubsection{The language associated with a formula}

Let now $\varphi(x_1,\ldots ,x_r,X_1,\ldots ,X_s)$ be a formula in which the free first order (resp. set) variables are $x_1,\ldots ,x_r$ (resp. $X_1,\ldots ,X_s$), with $r\le p$ and $s\le q$.

We interpret
\begin{itemize}
  \item $R_{a}$ as $R_a = \{i \in \Dom(u) \mid u_0(i) = a\}$;
  
  \item $x_{i}$ as the unique position of $1$ in $u_{i}$ (if $u_i\ne\epsilon$);
  
  \item $X_{j}$ as the set of positions of $1$ in $u_{p+j}$.
\end{itemize}
Note that if $p = q = 0$, then $\phi$ is a sentence and this is the usual notion of interpretation.

More formally, let $(u_0,u_1,\ldots,u_{p+q})$ be a non-empty word in $K_{p,q}$.  Let $n_i$ be the position of the unique $1$ in the word $u_i$ and let $N_j$ be the set of the positions of the 1's in the word $u_{p+j}$. We say that $u = (u_0,u_1,\ldots,u_{p+q})\in K_{p,q}$ satisfies $\phi$ if $u_0,\nu$ satisfy 
$\varphi$ where $\nu$ is the valuation defined by
$$\nu(x_i)=n_i\hbox{ for }1\leq i\leq r\quad\hbox{and}\quad\nu(X_j)=N_j\hbox{ for }1\leq  j\leq s.$$
We also say that the empty word (in $K_{p,q}$) satisfies $\phi$ if $\epsilon \models \phi$.
We let $L_{p,q}(\varphi) = \{u\in K_{p,q}\mid u \hbox{ satisfies }\varphi\}$.  Thus each \emph{formula} $\phi$ defines a subset of $K_{p,q}$, and hence a language in $B^*_{p,q}$.

\begin{example}
Let $\phi = \exists x\ (x<y \et R_{a}y)$. Then $FV(\phi) = \{y\}$. And $L_{1,0}(\phi)$ is the set of pairs of words $(u_{0},u_{1})$ such that $u_{0}\in A^*$, $u_{1}\in\{0,1\}^*$, 
$u_{0}$ and $u_{1}$ have the same length, $u_{1}$ has a single $1$, which is not the first position, and $u_{0}$ has an $a$ in that position.

Let $\phi = \forall x\ ((Xx\et x<y\et R_{b}y)\rightarrow R_{a}x)$. Then $L_{1,1}(\phi)$ is the set of triples of words $(u_{0},u_{1},u_{2})$ with $u_{0}\in A^*$, $u_{1},u_{2}\in \{0,1\}^*$, all three words have the same length, and either this length is zero, or $u_{1}$ has a single $1$ such that:

{\narrower Let $n$ be the position in $u_{1}$ which has a $1$. If $u_{0}$ has a $b$ in 
position $n$, then $u_{0}$ has an $a$ in each position before $n$ in which $u_{2}$ has a 
$1$. If $u_{0}$ does not have a $b$ in position $n$, then there is no constraint.\par}
\end{example}

\subsubsection{The $\MSO(<)$-definable languages are extended rational}

We first consider the languages associated with an atomic formula. Let $1\le i,j\le p+q$ and let $a\in A$. Let
\begin{align*}
C_{j,a} &= \{b \in B_{p,q} \mid b_j = 1 \hbox{ and }b_0 = a\}, \\
C_{i,j} &= \{b \in B_{p,q} \mid b_i = b_j = 1 \}, \\
\textrm{and }C_i     &= \{b \in B_{p,q} \mid b_i = 1 \}. \\
\end{align*}
Then we have
\begin{align*}
L_{p,q}(R_ax_i)  &= K_{p,q} \cap B_{p,q}^*C_{i,a}B_{p,q}^* \\
L_{p,q}(x_i=x_j) &= K_{p,q} \cap B_{p,q}^*C_{i,j}B_{p,q}^* \\
L_{p,q}(x_i<x_j) &= K_{p,q} \cap B_{p,q}^*C_iB_{p,q}^*C_jB_{p,q}^* \\
L_{p,q}(X_ix_j)  &= K_{p,q} \cap B_{p,q}^*C_{i+p,j}B_{p,q}^*. \\
\end{align*}
Thus, the languages defined by the atomic formulas, namely $L_{p,q}(R_ax)$, $L_{p,q}(x=y)$, $L_{p,q}(x < y)$ and $L_{p,q}(Xy)$, are extended rational.

Now let $\phi$ and $\psi$ be formulas and let us assume that $L_{p,q}(\phi)$ and $L_{p,q}(\psi)$ are extended rational. Then we have
\begin{align*}
L_{p,q}(\phi\vee\psi) &= L_{p,q}(\phi)\cup L_{p,q}(\psi)\\
L_{p,q}(\phi\wedge\psi) &= L_{p,q}(\phi)\cap L_{p,q}(\psi)\\
L_{p,q}(\neg\phi) &= K_{p,q}\setminus L_{p,q}(\phi),
\end{align*}
and hence these three languages are extended rational as well. We still need to handle existential quantification.

Let $\pi_i$ be the morphism which deletes the $i$-th component in a word of $B_{p,q}^*$;  that is: if $1\le i\le p$, then $\pi_{i}\colon B_{p,q}^* \rightarrow B_{p-1,q}^*$, and if $p< i\le p+q$, then $\pi_{i}\colon B_{p,q}^* \rightarrow B_{p,q-1}^*$.  In either case, we  have $\pi_{i}(b_0,b_1, \ldots, b_{p+q}) = (b_0,b_1,\ldots,b_{i-1},b_{i+1}, \ldots, 
b_{p+q})$.

Now, observe that, for any formula $\phi(x_1,\ldots ,x_r,X_1,\ldots ,X_s)$, and for $p\ge r$, $q\ge s$, $1\le i\le p$ and $1\le j\le q$ we have
$$L_{p-1,q}(\exists x_i\varphi) = \pi_i(L_{p,q}(\varphi))\hbox{\quad and \quad} L_{p,q-1}(\exists X_j\varphi) = \pi_{p+j}(L_{p,q}(\varphi)).$$
This concludes the proof that $L_{p,q}(\phi)$ is extended rational for any $p\ge r$, $q\ge s$.

In particular, if $\phi$ is a sentence in $\MSO(<)$ (that is, $\phi$ has no free variables), we may take $p = q = 0$. Then $L_{0,0}(\phi)$ is extended rational -- and we already noted that $L(\phi) = L_{0,0}(\phi)$.

\subsection{From extended rational expressions to automata}

It is immediately verified that the languages $\emptyset$, $\{\epsilon\}$, $\{a\}$ ($a\in A$) are accepted by finite automata. We now need to show that if $K,L \subseteq A^*$ are recognizable and if $\pi\colon A^*\rightarrow B^*$ is a morphism, then $\overline L$, $K\cup L$, $K\cap L$, $KL$, $K^*$ and $\pi(L)$ are recognizable.

\begin{proposition}\label{prop.complement}
	If $L\subseteq A^*$ is recognizable, then the complement $\overline L$ of $L$ is recognizable as well.
\end{proposition}

\begin{proof}
Let $\calA = (Q,\delta,i,F)$ be a deterministic complete automaton recognizing $L$. Then $\overline{\calA}=(Q,\delta,i,\overline{F})$ recognizes $\overline{L}$ by Proposition~\ref{prop: transition function}.
\eop

\begin{example}
The deterministic automata in Examples~\ref{A*abA* deterministe} and~\ref{exemple completion} confirm that, if $A = \{a,b\}$, then $b^*a^*$ is the complement of $A^*abA^*$.
\end{example}

Note that the resulting procedure yields a deterministic automaton for $\overline L$. It is very efficient if $L$ is given by a deterministic automaton, but may lead to an exponential growth in the number of states if $L$ is given by a non-deterministic automaton.

\begin{proposition}\label{prop.union and intersection}
	If $K,L \subseteq A^*$ are recognizable, then $K\cup L$ and $K\cap 	L$ are recognizable as well.
\end{proposition}

\begin{proof}
Let $\calA = \auto$ and $\calA' = \autop$ be automata recognizing $L$ and $L'$, respectively. We assume that the state sets $Q$ and $Q'$ are disjoint. Then it is readily verified that the automaton
$$\calA \cup \calA' = (Q \cup Q', T \cup T', I \cup I', F \cup F')$$
accepts $L \cup L'$. Thus $L\cup L'$ is recognizable, and hence so is $L \cap L' = \overline{\overline{L} \cup \overline{L'}}$, by Proposition~\ref{prop.complement}.
\eop

The construction in the above proof always yields a non-deterministic automaton for $L\cup L'$, even if we start from deterministic automata for $L$ and $L'$. The product of automata provides an alternative construction which preserves determinism, avoids any exponentiation of the number of states, and works for both the union and the intersection.

Let $\calA = \auto$ and $\calA' = \autop$ be automata recognizing the languages $L$ and $L'$. Their \emph{cartesian product} is the automaton $\calA'' = (Q \times Q',T'',I \times I',F\times F')$ where
$$T'' = \{((p,p'),a,(q,q')) \mid (p,a,q) \in T \hbox{ and } (p',a,q') \in T'\}.$$
Note that if $\calA$ and $\calA'$ are deterministic, then $\calA''$ is deterministic as well. The main property of $\calA''$ is the following: there exists a path $(p,p') \trans{u} (q,q')$ in $\calA''$ if and only if there exist paths $p \trans{u} q$ and $p' \trans{u} q'$, in $\calA$ and $\calA'$ respectively. Therefore $\calA''$ recognizes $L \cap L'$.

If we take $(F \times Q') \cup (Q \times F')$ as the set of final states, instead of $F \times F'$, and if  the
automata $\calA$ and $\calA'$ are complete, then the product automaton recognizes
$L \cup L'$.

In practice, the cartesian product of $\calA$ and $\calA'$ may not be trim, and one may want to use the procedure in Section~\ref{sec: trim} to produce more concise automata for $L \cap L'$ and $L \cup L'$.

\begin{remark}\label{addl rk 2}
Let us record here an algorithmic consequence of Propositions~\ref{prop.complement} and~\ref{prop.union and intersection}: given two automata $\calA$ and $\calB$, it is decidable whether $L(\calA) \subseteq L(\calB)$ and whether $L(\calA) = L(\calB)$. Indeed, we can compute automata accepting $L(\calA) \setminus L(\calB) = L(\calA) \cap \overline{L(\calB)}$ and $L(\calB) \setminus L(\calA)$, and decide whether these languages are empty (see Remark~\ref{addl rk 1}).
\end{remark}

\begin{proposition}
If $L, L' \subseteq A^*$ are recognizable, then $LL'$ and $L^*$ are recognizable as well.
\end{proposition}

\sketch
Let $\calA = \auto$ and let $\calA' = \autop$ be automata accepting $L$ and $L'$, respectively, and let us assume that their state sets are disjoint.

It is easily verified that the $\epsilon$-automaton
$$\big(Q \cup Q', T \cup T' \cup (F \times \{\epsilon\} \times I'), I, F'\big)$$

\pagebreak

\noindent
accepts $LL'$ (see Section~\ref{sec: epsilon}). Similarly, if $j$ is a state not in $Q$, the $\epsilon$-automaton
$$\big(Q \cup \{j\}, T \cup (F\times\{\epsilon\}\times I), I \cup\{j\}, F\cup\{j\}\big)$$
accepts $L^*$.
\eosketch

\begin{proposition}\label{prop.morphism}
If $L \subseteq A^*$ is recognizable and $\phi\colon A^* \to B^*$ is a morphism, then $\phi(L)$ is recognizable as well.
\end{proposition}

\sketch
Let $\calA = \auto$ be an automaton recognizing $L$.  We let $\calA'$ be the $\epsilon$-automaton $\calA' = (Q \cup Q',T',I,F)$, where the set $T'$ consists of
\begin{itemize}
\item[-] the transitions of the form $(p,\epsilon,q)$ such that $(p,a,q)\in T$ for some letter $a$ with $\phi(a) = \epsilon$,

\item[-] the transitions occurring in the paths of the form
$$p \trans{b_1} q'_1 \trans{b_2} \cdots\ q'_{k-1} \trans{b_k} q$$
such that $(p,a,q) \in T$, $\phi(a) = b_1 \cdots b_k \ne \epsilon$ and $q'_1,\ldots,q'_{k-1}$ are  new  states that we adjoin for each such triple $(p,a,q)$.
\end{itemize}
The set $Q'$ contains all the new states that occur in the latter paths. It is elementary to verify that $\calA'$ recognizes $\phi(L)$.
\eosketch

So far, we have shown that a language is recognizable, if and only if it is defined by a sentence in $\MSO(<)$, if and only if it is extended rational.

\begin{remark}\label{addl rk 3}
Note that the proofs of this logical equivalence are constructive, in the sense that given a sentence $\phi$ in $\MSO(<)$, we can construct an automaton $\calA$ such that $L(\phi) = L(\calA)$. It follows that $\MSO(<)$ is decidable: given an $\MSO$ sentence $\phi$, we can decide whether $\phi$ always holds. Indeed, this is the case if and only if $L(\neg\phi) = \emptyset$, which can be tested as discussed in Remark~\ref{addl rk 1}.
\end{remark}

\subsection{From automata to rational expressions}

To complete the proof of the Kleene-B\"uchi theorem, it suffices to prove that every recognizable language is rational. For this, we use the \emph{McNaughton-Yamada construction}\index{McNaughton-Yamada}.

Let $\calA=\auto$ be an automaton. For each pair of states $p,q\in Q$ and for each subset $P \subseteq 
Q$, let $L_{p,q}(P)$ be the set of all words $u\in A^*$ which label a 
path from state $p$ to state $q$, such that the states visited 
internally by that path are all in $P$:
\begin{align*}
   L_{p,q}(P) = \{ a_1a_2\ldots a_n \in A^* \mid
                &\textrm{ there exists a path in $\calA$} \\
   & p \trans{a_1} q_1
     \stackrel{a_2}{\longrightarrow} \ldots q_{n-1}
     \stackrel{a_n}{\longrightarrow} q
                    \hbox{ with } q_1,\ldots ,q_{n-1} \in P \}.
\end{align*}

Recall that, by convention, there always exists an empty path, 
labeled by the empty word, from any state $q$ to itself. So 
$\epsilon\in L_{p,q}(P)$ if and only if $p=q$.

We show by induction on the cardinality of $P$ that each language
$L_{p,q}(P)$ is rational. This will prove that $L(\calA)$ is rational, since $L(\calA) = \bigcup_{i\in I,\ f\in F}L_{i,f}(Q)$.

If $P=\emptyset$, then $L_{p,q}(\emptyset) = \{a\in A 
\mid (p,a,q)\in T\}$ if $p\ne q$, and $L_{q,q}(\emptyset) = \{a\in A 
\mid (q,a,q)\in T\}\cup\{\epsilon\}$. Thus $L_{p,q}(\emptyset)$ is 
always finite, and hence rational.

Now let $n>0$ and let us assume that, for any $p,q\in Q$ and $P\subseteq Q$ containing at most $n-1$ states, the language $L_{p,q}(P)$ is rational.  Let now $P\subseteq Q$ be a subset with $n$ elements and let $r\in P$. Considering the first and the last visit to state $r$ of a path from $p$ to $q$, we find that
$$L_{p,q}(P) = L_{p,q}(P\setminus \{r\})\enspace \cup \enspace
             L_{p,r}(P\setminus \{r\})
             L_{r,r}(P\setminus \{r\})^* L_{r,q}(P\setminus \{r\}).$$
Since $P\setminus \{r\}$ has cardinality $n-1$, it follows from the 
induction hypothesis that $L_{p,q}(P)$ is rational.

This concludes  the proof of the Kleene-B\"uchi theorem.

\subsection{Closure properties}\label{sec: closure ppties}

Rational languages enjoy many additional closure properties.

\begin{proposition}\label{prop.morph.inv}
Let  $\phi\colon A^{*}\rightarrow B^{*}$ be a morphism and let $L\subseteq B^{*}$. If $L$ is rational, then $\phi\inv(L)$ is rational as well.
\end{proposition}

\sketch
Let $\calA=\auto$ be an automaton over $B$, recognizing $L$, and let $\calA'=(Q,T',I,F)$ be the automaton over $A$ where
$$T' = \{(p,a,q) \mid p \trans{\phi(a)} q \textrm{ is a path in } \calA \}.$$
It is readily verified that $\calA'$ recognizes $\phi\inv(L)$.
\eosketch

Let $u\in A^*$ and $L\subseteq A^*$. The  \emph{left} and \emph{right quotients}\index{quotient} of $L$ by $u$ are defined as follows:
\begin{align*}
	u^{-1} L &= \{v \in A^* \mid uv \in L\}; \\
	L u^{-1} &= \{v \in A^* \mid vu \in L\}.
\end{align*}
These notions are generalized to languages: if $K$ and $L$ are 
languages, the \emph{left} and \emph{right quotients} of $L$ by $K$ are defined as follows:
\begin{align*}
	K^{-1} L &= \{v \in A^* \mid \exists u \in K \textrm{ such that } uv \in L\}
	          = \bigcup_{u \in K} u^{-1} L, \\
	L K^{-1} &= \{v \in A^* \mid \exists u \in K \textrm{ such that } vu \in L\}
	          = \bigcup_{u \in K} L u^{-1}.
\end{align*}

\begin{proposition}\label{prop.quotients}
If $L \subseteq A^*$ is rational and $K \subseteq A^*$ is any language (possibly not rational), then $K^{-1}L$ and $LK^{-1}$ are rational as well.
\end{proposition}

\sketch
If $\calA = (Q,T,I,F)$ is an automaton recognizing $L$. Let $I'$ be the set of states of $\calA$ which are accessible from an initial state of $\calA$ following a path labeled by a word of $K$,
$$I' = \{q \in Q \mid \exists i \in I, \exists u \in K \textrm{ such that } i \trans{u} q \}.$$
Then one shows that $\calA'=(Q,T,I',F)$ recognizes $K^{-1}L$. The proof for $LK^{-1}$ is similar.
\eosketch

\begin{remark}
The proof of Proposition~\ref{prop.quotients} is not effective: we may not be able to construct the set of states $I'$ associated with $K$. However, if $K$ is rational too, then $I'$ is effectively constructible.
\end{remark}

Recall that a word $u$ is a \emph{prefix}\index{prefix} of the word $v$ if there exists a word $v'\in A^{*}$ such that $v = uv'$ (that is: $v$ ``starts'' with $u$).  Similarly,  $u$ is a \emph{suffix}\index{suffix} of $v$ if there exists a word $v'\in A^{*}$ such that $v = v'u$. Finally  $u$ is a  \emph{factor}\index{factor} of $v$ if there exist words $v',v''\in A^{*}$ such that $v = v'uv''$.

If $L$ is a language, we let $\FG(L)$ (resp. $\FD(L)$, $\F(L)$) be the set of all prefixes (resp. suffixes, factors) of the words in $L$.

\begin{proposition}
If $L \subseteq A^*$ is rational, then $\FG(L)$, $\FD(L)$ and $\F(L)$ are rational as well.
\end{proposition}

\begin{proof}
The result follows from Proposition~\ref{prop.quotients}, since $\FG(L) = L(A^*)^{-1}$, $\FD(L) = (A^*)^{-1}L$ and $\F(L) = (A^*)^{-1}L(A^*)^{-1}$.
\eop

We leave it to the reader to verify that the following operations also preserve rationality.

The \emph{mirror image}\index{mirror image} of a word $u=a_1 \ldots a_n \in A^*$ is the word $\tilde{u} = a_n \ldots a_1$.  The corresponding language operation is given by $\tilde{L} = \{\tilde{u} \mid u \in L\}$ for each $L\subseteq A^*$.

A word $u=a_1 \ldots a_n \in A^*$ is a \emph{subword}\index{subword} of a word $v\in A^*$ if there exist words $u_0,\ldots,u_n \in A^*$ such that $v = u_0 a_1 u_1 \ldots a_n u_n$. If $L\subseteq A^*$, we let $\SM(L)$ be the set of all subwords of the words of $L$.

The \emph{shuffle}\index{shuffle} of the words $u$ and $v$ is the set
\begin{align*}
u \shuf v = \{w\in A^* \mid &\exists u_1,v_1,\ldots,u_n,v_n \in A^* \textrm{ such that } \\
	           &u = u_1 \cdots u_n,\ v = v_1 \cdots v_n \textrm{ and } w = u_1v_1 \cdots u_nv_n \}.
\end{align*}
If $K$ and $L$ are languages, we let $K \shuf L = \bigcup_{u\in K,\ v\in L} u \shuf v$.

\begin{proposition}
Let $K,\ L\subseteq A^*$ be rational languages. Then $\tilde L$, $\SM(L)$ and $K \shuf L$ are rational as well.
\end{proposition}

\section{Pumping lemmas}\label{section:pumping}

The characterizations summarized in the Kleene-B\"uchi theorem are sufficient most of the time to show that a language is rational. Showing  that a language is \emph{not} rational is a trickier problem. This short section presents the main tool for that purpose, namely the \emph{pumping lemma}\index{pumping lemma}. We actually first present a rather abstract version of this statement, and then its more classical corollaries.

\begin{theorem}\label{lem.iteration.3}
Let $L$ be a rational language.  There exists an integer $N>0$ 
with the following property.  For each word $w \in L$ and for each 
sequence of integers $0\le i_{0} < i_{1} < \ldots < i_{N} \le 
|w|$, there exist $0\le j < k\le N$ such that, if $w = u_1u_2u_3$ with $|u_{1}| = i_{j}$ and $|u_{1}u_{2}| = i_{k}$, then $u_1 u_2^* u_3 \subseteq L$.
\end{theorem}

\begin{proof}
Let $\calA$ be an automaton recognizing $L$, and let $N$ be the number of states of $\calA$.  Let $w=a_1 a_2 \cdots a_n\in L$ and let
$$p_0 \trans{a_1} p_1 \trans{a_2} p_2 \cdots \trans{a_n} p_n$$
be a successful path in $\calA$ labeled $w$.  Let $0 \leq i_0 < i_1 < 
\cdots < i_N \leq n$ be a sequence of integers.  Then two of the 
states $p_{i_0}, p_{i_1}, \ldots , p_{i_N}$ are equal, that is, there exist 
$0 \leq j < k \leq N$ such that  $p_{i_{j}}=p_{i_{k}}$.

Let $u_1 = a_1 \cdots a_{i_j}$, $u_2 = a_{1+i_j} \cdots a_{i_k}$ and $u_3 = a_{1+i_k} \cdots a_n$.  Of course, $w=u_1u_2u_3$, 
$|u_1|=i_j$, $|u_1u_2|=i_k$. The situation is summarized by Figure~\ref{fig: pumping}: we may iterate or skip the loop labeled $u_{2}$ and still retain a successful path, so $u_{1}u_{2}^{*}u_{3}\subseteq L$.
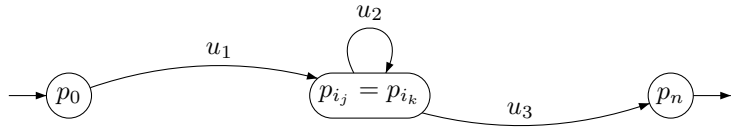
\begin{figure}[t]
\vspace*{5pt}
\begin{center}
\begin{picture}(90,12)(-5,-2)

\node[Nmarks=i,Nw=6.0,Nh=6.0](A)(0,0){$p_0$}
\node[Nw=16.0,Nh=6.0](B)(40,0){$p_{i_j} = p_{i_k}$}
\node[Nmarks=f,Nw=6.0,Nh=6.0](C)(80,0){$p_{n}$}

\drawedge[curvedepth=4](A,B){$u_1$}
\drawloop[loopdiam=6](B){$u_2$}
\drawedge[curvedepth=-4](B,C){$u_3$}

\end{picture}
\end{center}
\caption{Proof of the pumping lemma}
\label{fig: pumping}
\end{figure}
\eop

\begin{corollary}\label{lem.iteration.1}
Let $L$ be a rational language. There exists an integer $N>0$	such that, for each word $w \in L$ with length $|w| \geq N$, we can 	factor $w$ in three parts, $w=u_1u_2u_3$, with $u_2\neq\varepsilon$ and $u_1 u_2^* u_3 \subseteq L$.
\end{corollary}

\begin{corollary}\label{lem.iteration.2}
Let $L$ be a rational language. There exists an integer $N>0$
such that, for each word $w \in L$ with length $|w| \geq N$, we can 
factor $w$ in three parts, $w=u_1u_2u_3$, with $u_2\neq\varepsilon$,
$|u_{1}u_{2}|\le N$ (resp. $|u_{2}u_{3}|\le N$) and $u_1 u_2^* u_3 \subseteq L$.
\end{corollary}

\sketch
To prove Corollary~\ref{lem.iteration.2}, we apply Theorem~\ref{lem.iteration.3} with $i_{j} = j$ (resp. $i_{j} = n-N+j$) for $0\le j \le N$. And to prove Corollary~\ref{lem.iteration.1}, we take 
any sequence.
\eosketch


\begin{example}
It is a classical application of Corollary~\ref{lem.iteration.1} that $\{a^nb^n \mid n\ge 0\}$ is not rational: for each $N > 0$, the word $a^Nb^N$ cannot be factored as $w = u_1u_2u_3$ with $u_2 \ne\epsilon$ and $u_1u_2^*u_3 \subseteq \{a^nb^n \mid n\ge 0\}$.

Corollary~\ref{lem.iteration.2} can be used to show that $\{u\in \{a,b\}^* \mid |u|_a = |u|_b\}$ is not rational (take again $a^Nb^N$); however, this language satisfies the necessary condition for rationality in Corollary~\ref{lem.iteration.1}, with $N = 2$.

Consider now the following language over the alphabet $\{a,b,c,d\}$
$$\{(ab)^n (cd)^n \mid n\geq 0\} \cup A^* \{aa,bb,cc,dd,ac\} A^*$$
It satisfies the necessary condition for rationality in Corollary~\ref{lem.iteration.2}, but it is not rational, as can be proved using Theorem~\ref{lem.iteration.3}.
\end{example}

However, the pumping lemma as stated here may not be enough to prove that a given language is not rational.  Let us say that a word \emph{contains a square} if it can be written in the form $uvvw$ with $v\ne\epsilon$. Then the language
$$\{udv\mid u,v \in \{a,b,c\}^* \textrm{ and either $u\neq v$,
or one of $u$ and $v$ contains a square}\}$$
satisfies the necessary condition for rationality in Theorem~\ref{lem.iteration.3} (for $N=4$). Yet it is not rational (the proof of that fact uses the existence of arbitrarily long words on the alphabet $\{a,b,c\}$ containing no square).

Ehrenfeucht, Parikh, Rozenberg gave a necessary and sufficient condition for 
rationality in the same style as the pumping lemma (see \textit{e.g.} \cite[Theorem I.3.3]{Sakarovitch:2009}).

\section{Minimal automaton and syntactic monoid}\label{section:minimal}


\begin{figure}[t]
\begin{center}
\begin{picture}(110,49)(0,0)

\node[Nmarks=i,Nw=6.0,Nh=6.0](A)(0,22.5){$\epsilon$}
\node[Nw=6.0,Nh=6.0](B)(15,37.5){$a$}
\node[Nw=6.0,Nh=6.0](C)(15,7.5){$b$}
\node[Nw=6.0,Nh=6.0](D)(30,45){$aa$}
\node[Nw=6.0,Nh=6.0](E)(30,30){$ab$}
\node[Nw=6.0,Nh=6.0](F)(30,15){$ba$}
\node[Nw=6.0,Nh=6.0](G)(30,0){$bb$}
\node[Nmarks=f,Nw=6.0,Nh=6.0](H)(45,45){}

\drawedge(A,B){$a$}
\drawedge[ELside=r](A,C){$b$}
\drawedge(B,D){$a$}
\drawedge[ELside=r](B,E){$b$}
\drawedge(C,F){$a$}
\drawedge[ELside=r](C,G){$b$}
\drawedge(D,H){$a$}
\drawedge(D,E){$b$}
\drawedge[ELside=r](G,F){$a$}
\drawedge[curvedepth=2](E,F){$a$}
\drawedge[curvedepth=8](E,G){$b$}
\drawedge[curvedepth=2](F,E){$b$}
\drawloop[loopangle=-10](G){$b$}
\drawedge[ELside=r,curvedepth=-8](F,D){$a$}
\drawloop[loopangle=-70](H){$a,b$}

\node[Nmarks=i,Nw=6.0,Nh=6.0](A1)(60,15){$\epsilon$}
\node[Nw=6.0,Nh=6.0](B1)(75,15){$a$}
\node[Nw=6.0,Nh=6.0](C1)(90,15){$aa$}
\node[Nmarks=f,Nw=8.0,Nh=6.0](D1)(105,15){$aaa$}

\drawloop(A1){$b$}
\drawedge[curvedepth=2](A1,B1){$a$}
\drawedge[curvedepth=2](B1,A1){$b$}
\drawedge(B1,C1){$a$}
\drawedge[curvedepth=10](C1,A1){$b$}
\drawedge(C1,D1){$a$}
\drawloop(D1){$a,b$}

\end{picture}
\end{center}
\caption{Two different automata for $A^*aaaA^*$}
\label{fig: minimal vs. non minimal automaton}
\end{figure}
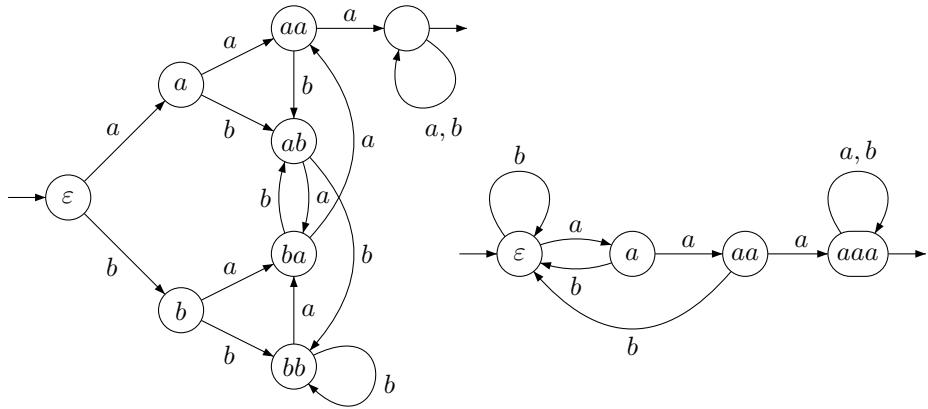

Consider the two automata in Figure~\ref{fig: minimal vs. non minimal automaton}. Both are complete and deterministic, and both recognize the set of words over $A=\{a,b\}$ that contain some occurrence of the word $aaa$ as a factor --- that is, the language  $A^*aaaA^*$. The two automata were designed using different intuitions about how to go about this task:  In the first instance, the underlying algorithm is ``keep track of the last two letters read from the input'', as indicated by the state labels, while in the second automaton the algorithm is, ``keep track of the length of the longest suffix of $a$'s in the input''.  Thus the second automaton achieves the same result with a smaller number of states.  It is easy to see that the second example is also optimal---no complete deterministic automaton recognizing this language can have a smaller number of states.

In this section we will see that for every rational language $L$ there is a unique minimal complete deterministic automaton accepting $L$.  We will also describe an efficient algorithm that takes as input an arbitrary complete deterministic automaton $\calA$, and produces as output the minimal automaton for $L(\calA)$.

\subsection{Myhill-Nerode equivalence and the minimal automaton}

One way to see that there is something inefficient about the first automaton in the example above is to observe its behavior on the two input words $u=bab$ and $v=abb$.  These words lead from the initial state to two different states. However, for purposes of recognizing words in $L$, there is no point in distinguishing between $u$ and $v$, for no matter what the subsequent input $w$ is, the result will be the same:  either $uw$ and $vw$ are both in $L$ or both outside of $L$. 

To formalize this notion of inputs that are indistinguishable with respect to $L$, we make the following definitions:  If $u,v\in A^*$ we define $u\equiv_L v$ if and only if $u^{-1}L=v^{-1}L$ (see Section~\ref{sec: closure ppties}). Obviously, $\equiv_L$ is an equivalence relation on $A^*$.  We also note that if $u\equiv_L v$, and $w\in A^*$, then $uw\equiv_L vw$, since $(uw)^{-1}L = w^{-1}(u^{-1}L)$.  An equivalence relation with this multiplicative property is said to be a \emph{right congruence}\index{congruence!right}\index{Myhill-Nerode congruence}.  Further, $L$ itself is a union of $\equiv_L$-classes, since $w\in L$ if and only if $\epsilon\in w^{-1}L$.

We can accordingly define a complete deterministic automaton $\calA_{\rm min}(L)$ by making the states these classes of equivalent words:  We set $\calA_{\rm min}(L) = (Q_L,\delta_L,i_L,F_L)$, where $Q_L=A^*/\equiv_L$, $i_L=[\epsilon]_{\equiv_L}$, and $F_L$ and $\delta_L\colon Q_L\times A\to Q_L$ are defined by
$$F_L=\{[v]_{\equiv_L} \mid v\in L\} \qquad\textrm{and}\qquad\delta([v]_{\equiv_L},a)=[va]_{\equiv_L}.$$
We need to show that  this is well-defined, since a state will in general have many different representations of the form $[v]_{\equiv_L}$.  But well-definedness is an immediate consequence of our observation that $\equiv_L$ is a right congruence.
We have the following result.

\begin{theorem}\label{thm:myhill_nerode}
Let $L\subseteq A^*$.
\begin{enumerate}
\item $\calA_{\rm min}(L)$ accepts $L$.
\item $L$ is rational if and only if $\equiv_L$ has finite index.
\end{enumerate}
\end{theorem}

\begin{proof}
It follows at once by induction on $|w|$ that for all $w\in A^*$,
$$\delta_L([\epsilon]_{\equiv_L},w)=[w]_{\equiv_L}.$$
Since, as observed above, $L$ itself is a union of $\equiv_L$-classes, it follows that $w$ is accepted if and only if $w\in L$.  This proves the first claim.

To prove the second claim in the theorem, note that if $\equiv_L$ has finite index, then $\calA_{\rm min}$ is a finite automaton, and therefore by {\it (1),} $L$ is rational.  Conversely, if $L$ is rational, then it is accepted by some complete deterministic automaton $(Q,\delta,i,F)$ with $Q$ finite.  Now suppose $u, v\in A^*$ and $\delta(i,u)=\delta(i,v)$.  Then if $w\in A^*$ and $uw\in L$, we have
$$\delta(i,vw) = \delta(\delta(i,v),w) = \delta(\delta(i,u),w) = \delta(i,uw)\in F,$$
so $vw\in L$.  Similarly, $vw\in L$ implies $uw\in L$, so $u \equiv_L v$.  Thus the number of classes of $\equiv_L$ cannot be more than $|Q|$, so $\equiv_L$ has finite index.
\eop

The proof of Theorem~\ref{thm:myhill_nerode} shows that $\calA_{\rm min}(L)$ has the least number of states among the complete deterministic automata accepting $L$. The automaton $\calA_{\rm min}(L)$ is called the \emph{minimal automaton}\index{automaton!minimal} of $L$. We now give another, more algebraic justification for this terminology.

\subsection{Uniqueness and minimality of $\calA_{\rm min}(L)$}

Let $\calA = (Q,\delta,i,F)$ be a complete deterministic automaton over $A$, and let $L=L(\calA)$.  We say that $p,q\in Q$ are \emph{equivalent}\index{state!equivalent} states, and write $p\equiv q$, if
$$\{v\in A^* \mid \delta(p,v)\in F\} = \{v\in A^* \mid \delta(q,v)\in F\}.$$
Intuitively, this means that for purposes of recognizing words in $L$, $p$ and $q$ do the same job, and we might as well merge them into a single state.

We now repeat, in a somewhat different form, an observation made in the proof of Theorem~\ref{thm:myhill_nerode}:  If $\delta(i,u)\equiv \delta(i,v)$, then
$$uw \in L \Longleftrightarrow \delta(\delta(i,u),w)\in L   \Longleftrightarrow \delta(\delta(i,v),w)\in L  \Longleftrightarrow vw\in L,$$
so that $u\equiv_L v$.  In particular, if $\delta(i,u)=\delta(i,v)$, then $u\equiv_L v$, so we have a well-defined mapping $\delta(i,w)\mapsto [w]_{\equiv_L}$, from the set of {\it accessible} states of $\calA$ onto the states of $\calA_{\rm min}(L)$.  Note that this mapping sends the initial state $i=\delta(i,\epsilon)$ to $[\epsilon]_{\equiv_L}$, final states of $\calA$ to final states of $\calA_{\rm min}(L)$, and respects the next-state function.  We summarize these observations as follows.

\begin{theorem}\label{thm:minimal_automaton}
Let $\calA=(Q,\delta,i,F)$ be a complete deterministic automaton over $A$, and let $L=L(\calA)$.  Then there is a map $f$ from the set of accessible states in $Q$ onto $Q_L$ such that
\begin{itemize}
\item for all $a\in A$ and accessible $q\in Q$, $f(\delta(q,a))=\delta_L(f(q),a)$,
\item $f(i)=i_L$,
\item $f(F)=F_L$.
\end{itemize}
Moveover, $f(p)=f(q)$ if and only if $p\equiv q$.
\end{theorem}
In particular, if $\calA$ has the same number of states as $\calA_{\rm min}(L)$, then since $f$ is onto, the two automata are isomorphic by Theorem~\ref{thm:minimal_automaton}.

\subsection{An algorithm for computing the minimal automaton}

Theorem~\ref{thm:minimal_automaton} says that in principle we can compute the minimal automaton of a rational language $L$ starting from any complete deterministic automaton $(Q,\delta,i,F)$ accepting $L$, first by removing the inaccessible states and then merging equivalent states.  We have already seen how to compute the accessible states. How do we determine if two states are equivalent?  If $p,q$ are inequivalent states then there is a word $v\in A^*$ that distinguishes between these states in the sense  that $\delta(p,v)\in F$ and $\delta(q,v)\notin F$, or vice-versa.  It follows from a simple pumping argument that if such a distinguishing word exists, then it can be chosen to have length no more than $|Q|^2$.  Thus we can effectively determine whether two states are equivalent by calculating $\delta(p,v)$  and $\delta(q,v)$ for all words up to this length.  

Of course, this is a terrible algorithm, since there are $|A|^{|Q|^2}$ different words to check!  In practice, we can proceed as follows:  Let $m\geq 0$.  We say $p \equiv_m q$ if for all $v\in A^*$ of length no more than $m$, $\delta(p,v)\in F$ if and only if $\delta(q,v)\in F$.  This is clearly an equivalence relation on $A^*$, and $\equiv_{m+1}$ refines $\equiv_m$ for all $m$.  The following lemma improves the $|Q|^2$ bound on the length of distinguishing words.

\begin{lemma}\label{lemma:state_equivalence}
Let $p,q\in Q$.  Then $p\equiv q$ if and only if $p\equiv_m q$ for $m=|Q|-2$.
\end{lemma}

\begin{proof}
First suppose that for some $m$, the equivalence relations $\equiv_m$ and $\equiv_{m+1}$ coincide.  We claim that $\equiv_m$ and $\equiv$ coincide. To see this, suppose that $p$ and $q$ are inequivalent, and that $w$ is a word of minimal length distinguishing them.  If $|w| > m$, then we can write $w = uv$, where $|v|=m+1$, so that $p'=\delta(p,u)$ and $q'=\delta(q,u)$ are inequivalent modulo $\equiv_{m+1}$.  But this means that they are also inequivalent modulo $\equiv_m$, and thus distinguished by a word $v'$ of length no more than $m$, and thus $p$ and $q$ are distinguished by the word $uv'$ of length strictly less than that of $w$, a contradiction.  Thus the minimal distinguishing word has length no more than $m$, so that $\equiv_m$ coincides with $\equiv$. 

Now if $\equiv_{m+1}$ does not coincide with $\equiv_m$, then $\equiv_{m+1}$ has a larger number of classes.  Since the number of classes can never exceed $|Q|$, and since $\equiv_0$ has two classes, the sequence $\{\equiv_m\}_{m\geq 0}$ will stabilize by the time $m$ reaches $|Q|-2$.
\eop

Lemma~\ref{lemma:state_equivalence} leads to the following practical algorithm for minimization.  We begin with a list of all the pairs $\{p,q\}$ of distinct accessible states, and mark the pair if $p\in F$ and $q\notin F$, or vice-versa. In each phase of the algorithm, we visit each unmarked pair $\{p,q\}$ and each $a\in A$, we compute $\{p',q'\}= \{\delta(p,a),\delta(q,a)\}$, and we mark $\{p,q\}$ if $\{p',q'\}$ is marked.  An easy induction shows that if a pair $\{p,q\}$ is distinguished by a word of length $m$, then it will be marked by the $m^{th}$ phase of the algorithm.  Thus after no more than $|Q|-2$ phases, the algorithm will not mark any new pairs, with the result that the algorithm terminates, and the unmarked pairs are exactly the pairs of equivalent states.

\begin{example}
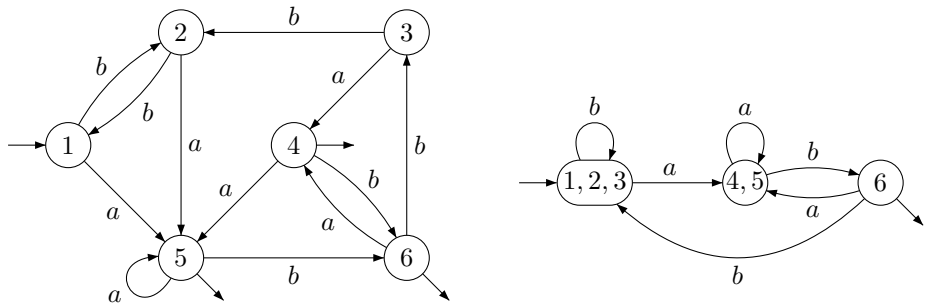
\begin{figure}[t]
\vspace*{-36pt}
\begin{center}
\begin{picture}(110,49)(0,-3)

\node[Nmarks=i,Nw=6.0,Nh=6.0](A)(0,15){$1$}
\node[Nw=6.0,Nh=6.0](B)(15,30){$2$}
\node[Nw=6.0,Nh=6.0](C)(45,30){$3$}
\node[Nmarks=f,Nw=6.0,Nh=6.0](D)(30,15){$4$}
\node[fangle=-45,Nmarks=f,Nw=6.0,Nh=6.0](E)(15,0){$5$}
\node[fangle=-45,Nmarks=f,Nw=6.0,Nh=6.0](F)(45,0){$6$}

\drawedge[ELside=r](A,E){$a$}
\drawedge[curvedepth=2](A,B){$b$}
\drawedge(B,E){$a$}
\drawedge[curvedepth=2](B,A){$b$}
\drawedge[ELside=r](C,D){$a$}
\drawedge[ELside=r](C,B){$b$}
\drawedge[ELside=r](D,E){$a$}
\drawedge[curvedepth=2](D,F){$b$}
\drawloop[loopdiam=5,loopangle=-150](E){$a$}
\drawedge[ELside=r](E,F){$b$}
\drawedge[curvedepth=2](F,D){$a$}
\drawedge[ELside=r](F,C){$b$}

\node[Nmarks=i,Nw=10.0,Nh=6.0](A1)(70,10){$1,2,3$}
\node[Nw=6.0,Nh=6.0](B1)(90,10){$4,5$}
\node[fangle=-45,Nmarks=f,Nw=6.0,Nh=6.0](C1)(108,10){$6$}

\drawedge[loopdiam=5](A1,B1){$a$}
\drawloop[loopdiam=5](A1){$b$}
\drawloop[loopdiam=5](B1){$a$}
\drawedge[curvedepth=2](B1,C1){$b$}
\drawedge[curvedepth=2](C1,B1){$a$}
\drawedge[curvedepth=10](C1,A1){$b$}

\end{picture}
\end{center}
\caption{The minimization algorithm}
\label{fig: minimization}
\end{figure}
Consider the first automaton in Figure~\ref{fig: minimization}. Initially we mark the pairs $\{i,j\}$, where $i\in\{1,2,3\}$ and $j\in\{4,5,6\}$.  On the next pass, the pairs $\{4,6\}$ and $\{5,6\}$ are marked since applying $b$ to these pairs gives the marked pair $\{3,6\}$.  No further pairs are marked on the next pass, so the algorithm terminates. Since the pairs $\{1,2\}$ and $\{2,3\}$ are unmarked, $\{1,2,3\}$ is an equivalence class, and since $\{4,5\}$ is unmarked, it forms a second class.  The remaining class is $\{6\}$.  The resulting minimal automaton is pictured on the right-hand side of Figure~\ref{fig: minimization}.
\end{example}

\begin{example} We now apply the algorithm to the automaton in 
Figure~\ref{fig: already minimal}. Initially, the pairs $\{i,6\}$ with $i<6$ are marked. 
On the next pass the pairs $\{i,5\}$ with $i<5$ are marked, {\it etc.,} until on the 
fifth pass the pair $\{1,2\}$ is marked. The result is that every pair of distinct 
states is marked: the automaton is already minimal.
\end{example}

\begin{figure}[t]
\begin{center}
\begin{picture}(105,9)(0,0)
\node[Nmarks=i,Nw=6.0,Nh=6.0](A)(0,0){$1$}
\node[Nw=6.0,Nh=6.0](B)(20,0){$2$}
\node[Nw=6.0,Nh=6.0](C)(40,0){$3$}
\node[Nw=6.0,Nh=6.0](D)(60,0){$4$}
\node[Nw=6.0,Nh=6.0](E)(80,0){$5$}
\node[fangle=-45,Nmarks=f,Nw=6.0,Nh=6.0](F)(100,0){$6$}

\drawedge(A,B){$a$}
\drawedge(B,C){$a$}
\drawedge(C,D){$a$}
\drawedge(D,E){$a$}
\drawedge(E,F){$a$}
\drawloop[loopdiam=5](A){$b$}
\drawloop[loopdiam=5](B){$b$}
\drawloop[loopdiam=5](C){$b$}
\drawloop[loopdiam=5](D){$b$}
\drawloop[loopdiam=5](E){$b$}
\drawloop[loopdiam=5](F){$a,b$}

\end{picture}
\end{center}
\caption{A minimal automaton}
\label{fig: already minimal}
\end{figure}
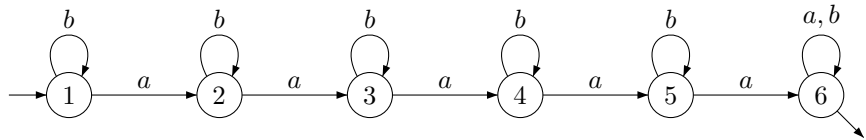

The pair-marking implementation of the algorithm just illustrated is suitable for small examples worked by hand.  In the worst case, shown in the last example, we check $\O(|Q|^2)$ unmarked pairs on each pass, and make $\O(|Q|)$ passes, with $|A|$ consultations of the state-transition table for each pair we inspect.  Thus, the overall time complexity of the algorithm is $\O(|A|\cdot |Q|^3)$.  More astute bookkeeping, in which we partition equivalence classes at each step, rather than marking pairs of inequivalent states, leads to a $\O(|A|\cdot |Q|^2)$ algorithm (Moore~\cite{Moore:1956}).  This can be further improved to $\O(|A|\cdot |Q|\cdot \log |Q|)$ (Hopcroft~\cite{Hopcroft:1971}).

\subsection{The transition monoid of an automaton}

Let $\calA=(Q,\delta,i,F)$ be a complete deterministic automaton over an alphabet $A$.  Let $w\in A^*$.  We study the maps
$$f_w^{\calA}\colon q\longmapsto \delta(q,w)$$
from $Q$ into itself.  We will write the image of a state $q$ under $f_w^{\calA}$ as $qf_w^{\calA}$ rather than the more traditional $f_w^{\calA}(q)$.  We then have, for $v,w\in A^*$,
$$f_{vw}^{\calA}=f_v^{\calA}f_w^{\calA},$$
where the product in the right-hand side of the equation is left-to-right composition of functions --- that is, $q(f_v^{\calA}f_w^{\calA})=(qf_v^{\calA})f_w^{\calA}$.  

We will henceforth drop the superscript $\calA$, except in situations where several different automata are involved.  Observe that $f_\epsilon$ is the identity map on $Q$.  Thus the set of maps 
$$M(\calA)=\{f_w \mid w\in A^*\}$$
forms an algebraic structure with an associative product and an identity element (usually denoted 1).  Such a structure is called a \emph{monoid}\index{monoid}, and we call $M(\calA)$ the \emph{transition monoid}\index{monoid!transition}\index{transition!monoid} of $\calA$.  Observe that if $Q$ is finite, then $M(\calA)$ is finite, and that the structure of $M(\calA)$ depends only on the next-state function $\delta$, and not at all on the initial or final states.

$A^*$ is, of course, itself a monoid, with concatenation of words as the operation and the empty word $\epsilon$ as the identity.  The map
$$\phi\colon w\longmapsto f_w$$
is consequently a monoid \emph{morphism} from $A^*$ into $M(\calA)$; that is, it satisfies
$$\phi(w_1w_2)=\phi(w_1)\phi(w_2)$$
for all $w_1,w_2$ in $A^*$, and it maps the identity element of $A^*$ to the identity element of $M(\calA)$.

\begin{example}~\label{ex:monoid-abstar}
In the diagrams in this example and in Examples~\ref{ex: A2} and~\ref{ex:fulltm}, we indicate only the transitions between states, since, as we have observed, the initial and final states do not enter into the computation of the transition monoid of an automaton.

First, consider the automaton $\calA_1$ in Figure~\ref{fig: monoid-abstar}.
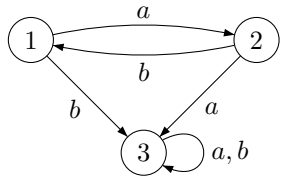
\begin{figure}[t]
\vspace*{5pt}
\begin{center}
\begin{picture}(34,18)(0,0)
\node[Nw=6.0,Nh=6.0](A)(0,15){$1$}
\node[Nw=6.0,Nh=6.0](B)(30,15){$2$}
\node[Nw=6.0,Nh=6.0](C)(15,0){$3$}

\drawedge[curvedepth=2](A,B){$a$}
\drawedge[curvedepth=2](B,A){$b$}
\drawedge[ELside=r](A,C){$b$}
\drawedge(B,C){$a$}
\drawloop[loopangle=0,loopdiam=5](C){$a,b$}
\end{picture}
\end{center}
\caption{The automaton $\calA_1$, with no indication of initial or terminal states}
\label{fig: monoid-abstar}
\end{figure}
We will write an element $f_w$ of $M(\calA_1)$ as a  vector $f_w = (\begin{array}{ccc} 1f_w & 2f_w & 3f_w\end{array})$.
We can then begin enumerating the elements of $M(\calA_1)$:
\begin{align*}
&f_\epsilon = (\begin{array}{ccc}1&2&3\end{array})\qquad\\
&f_a = (\begin{array}{ccc}2&3&3\end{array}) \qquad
f_b = (\begin{array}{ccc}3&1&3\end{array})\\
&f_{aa} = (\begin{array}{ccc}3&3&3\end{array})\qquad
f_{ab} = (\begin{array}{ccc}1&3&3\end{array})\qquad
f_{ba} = (\begin{array}{ccc}3&2&3\end{array})\qquad
f_{bb} = (\begin{array}{ccc}3&3&3\end{array})
\end{align*}
We could continue enumerating like this, but instead we note that $f_{aba}=f_a$, $f_{bab}=f_b$, and for all other $w\in A^*$ of length 3, $f_w=(\begin{array}{ccc}3&3&3\end{array})$.  Thus the inventory above is the entire transition monoid, since any transition induced by a word of length greater than 2 is equal to one induced by a shorter word.  Thus $M(\calA)$ has 6 elements $1,\alpha=f_a,\beta=f_b,\alpha\beta,\beta\alpha$, and 0.  The multiplication is then determined by the laws $\alpha\alpha=\beta\beta=0$, $\alpha=\alpha\beta\alpha$, and $\beta=\beta\alpha\beta$. The complete multiplication table is shown below:
\begin{center}
\begin{tabular}{|c||c|c|c|c|c|c|}
\hline
$\cdot$&1&$\alpha$&$\beta$&$\alpha\beta$&$\beta\alpha$&0\\
\hline\hline
1&1&$\alpha$&$\beta$&$\alpha\beta$&$\beta\alpha$&0\\
\hline
$\alpha$&$\alpha$&0&$\alpha\beta$&0&$\alpha$&0\\
\hline
$\beta$&$\beta$&$\beta\alpha$&0&$\beta$&0&0\\
\hline
$\alpha\beta$&$\alpha\beta$&$\alpha$&0&$\alpha\beta$&0&0\\
\hline
$\beta\alpha$&$\beta\alpha$&0&$\beta$&0&$\beta\alpha$&0\\
\hline
0&0&0&0&0&0&0\\
\hline
\end{tabular}
\end{center}
This example illustrates an important general point:  There is an effective procedure for computing the multiplication table of the transition monoid of a complete deterministic finite automaton.  We enumerate the maps $f_w$ until we find that all words of some length induce the same maps as shorter words.

\end{example}

\begin{example}\label{ex: A2}
Consider the automaton $\calA_2$ in Figure~\ref{fig: 5state}.
\begin{figure}[t]
\vspace*{5pt}
\begin{center}
\begin{picture}(90,30)(0,-6)
\node[Nw=5.0,Nh=5.0](A)(10,21){$1$}
\node[Nw=5.0,Nh=5.0](B)(25,21){$2$}
\node[Nw=5.0,Nh=5.0](C)(30,7){$3$}
\node[Nw=5.0,Nh=5.0](D)(17.5,0){$4$}
\node[Nw=5.0,Nh=5.0](E)(5,7){$5$}
\drawedge[curvedepth=2](A,B){$a,b$}
\drawedge[curvedepth=2](B,A){$b$}
\drawedge[ELside=r](B,C){$a$}
\drawedge[ELside=r](C,D){$a$}
\drawedge[ELside=r](D,E){$a$}
\drawedge[ELside=r](E,A){$a$}
\drawloop[loopangle=-30,loopdiam=5](C){$b$}
\drawloop[loopangle=-90,loopdiam=5](D){$b$}
\drawloop[loopangle=-150,loopdiam=5](E){$b$}
\put(-5,-6){$\calA_2$}

\node[Nw=5.0,Nh=5.0](A1)(65,21){$1$}
\node[Nw=5.0,Nh=5.0](B1)(80,21){$2$}
\node[Nw=5.0,Nh=5.0](C1)(85,7){$3$}
\node[Nw=5.0,Nh=5.0](D1)(72.5,0){$4$}
\node[Nw=5.0,Nh=5.0](E1)(60,7){$5$}
\drawedge[curvedepth=2](A1,B1){$a,b,c$}
\drawedge[curvedepth=2](B1,A1){$b$}
\drawedge[ELside=r](B1,C1){$a$}
\drawedge[ELside=r](C1,D1){$a$}
\drawedge[ELside=r](D1,E1){$a$}
\drawedge[ELside=r](E1,A1){$a$}
\drawloop[loopangle=0,loopdiam=5](B1){$c$}
\drawloop[loopangle=-30,loopdiam=5](C1){$b,c$}
\drawloop[loopangle=-90,loopdiam=5](D1){$b,c$}
\drawloop[loopangle=-150,loopdiam=5](E1){$b,c$}
\put(90,-6){$\calA_3$}
\end{picture}
\end{center}
\caption{The automata $\calA_2$ and $\calA_3$}
\label{fig: 5state}
\end{figure}
The transition monoid is generated by the two permutations $f_a$ and $f_b$, both of which are permutations of the set of states: $f_a$ cycles the five states and $f_b$ transposes a pair of adjacent states. It is well known from elementary group theory that we can obtain all transpositions $t$ of adjacent elements by repeated conjugation with the cycle (the map $t\mapsto f_a^4\,t\,f_a$),
and that all permutations of the states can be obtained by composing transpositions of pairs of adjacent elements.  So $M(\calA)$ consists of all the permutations of $\{1,2,3,4,5\}$, and is consequently the symmetric group of degree 5, with $5!=120$ elements.  Of course, we can do likewise with any finite set of states.
\end{example}

\begin{example}~\label{ex:fulltm}
Now consider the effect of adding a third input letter to the preceding example, obtaining the automaton $\calA_3$ in Figure~\ref{fig: 5state}. It is not hard to show that {\it every} map from $\{1,2,3,4,5\}$ into itself can be obtained by repeatedly composing $f_c$ with permutations.  Thus $M(\calA_3)$ is the \emph{full transformation monoid}\index{monoid!full transformation} on  5 states, which has $5^5=3125$ elements.  We can similarly generate a transition monoid with $n^n$ elements using an $n$-state automaton.
\end{example}

\subsection{The syntactic monoid}

Now let $L\subseteq A^*$, and consider the transition monoid of the minimal automaton
$\calA_{\rm min}(L)=(Q_L,\delta_L,i_L,F_L)$.  Let $u,v\in A^*$. When are the two elements $f_u$, $f_v$ of this monoid the same? If they are different, then there is some state $q$ such that $qf_u\neq qf_v$.  Since the automaton is minimal, there is a word $y\in A^*$ distinguishing these two states, so that $qf_uf_y\in F_L$ and $qf_vf_y\notin F_L$, or vice-versa.  Since every state is accessible, there is also a word $x$ such that $q=if_x$, so that either $xuy\in L$ and $xvy\notin L$, or vice-versa. Conversely, if such a pair of words $x,y$ exists, then $f_u$ and $f_v$ cannot be equal.  We thus have:

\begin{theorem}\label{thm:syntactic-congruence}
Let $L\subseteq A^*$, and let $u,v\in A^*$.  Let $\calA=\calA_{\rm min}(L)$.  Then $f_u^{\calA}=f_v^{\calA}$ if and only if for all $x,y\in A^*$
$$xuy\in L\Longleftrightarrow xvy\in L.$$
\end{theorem}

If the conditions in this theorem are satisfied, then we write $u\cong_L v$. The equivalence relation $\cong_L$ is called the \emph{syntactic congruence}\index{congruence!syntactic} of $L$, and the transition monoid of $\calA_{\rm min}(L)$ is called the \emph{syntactic monoid}\index{monoid!syntactic} of $L$.  We denote the syntactic monoid of $L$ by $M(L)$.  In algebraic terms, $M(L)=M(\mathcal{A}_\text{min}(L))=A^*/{\cong_L}$, that is $M(L)$ is the quotient monoid of $A^*$ by the syntactic congruence.  The morphism mapping each $w\in A^*$ to its $\cong_L$-class is called the \emph{syntactic morphism}\index{morphism!syntactic} of $L$, and is denoted $\mu_L$.

The syntactic congruence is a two-sided congruence on $A^*$; that is, if $u\cong_L v$ and $u'\cong_L v'$, then $uu'\cong_L vv'$.  Compare this to the Myhill-Nerode congruence $\equiv_L$, which, as we noted, is a  right congruence. The equivalence $\cong_L$ refines $\equiv_L$.

Transition monoids, and, in particular, the syntactic monoid, allow us to place many questions about the behavior of automata in a purely algebraic setting.  For instance, we have the following algebraic characterization of rationality:  Let $M$ be a monoid and $\phi \colon A^*\to M$ a morphism.  We say that $\phi$ \emph{recognizes}\index{recognition} $L\subseteq A^*$ if and only if there is a subset $X$ of $A^*$ such that $L=\phi^{-1}(X)$.  We also say in this situation that $M$ recognizes $L$.

\begin{theorem}\label{thm:rational-recognizable}
Let $L\subseteq A^*$.  The following are equivalent:
\begin{enumerate}
\item $L$ is rational.
\item $M(L)$ is finite.
\item $L$ is recognized by a finite monoid.
\end{enumerate}
\end{theorem}

\begin{proof}
To show {\it (1)} implies {\it (2)}, note that if $L$ is rational, then $\calA_{\rm min}(L)$ has a finite set of states, and thus its transition monoid, $M(L)$, is finite. For {\it (2)} implies {\it (3)}, if $u\in L$ and $u\cong_L v$, then $v=\epsilon\, v\, \epsilon$ is also in $L$.  Thus $L$ is a union of equivalence classes of $\cong_L$, so that $L=\mu_L^{-1}(X)$, where $X=\{f_w\in M(\calA_{\rm min}(L)) \mid w\in L\}$. Finally, to show {\it (3)} implies {\it (1)}, suppose $\phi\colon A^*\to M$, where $M$ is finite,  and that $L=\phi^{-1}(X)$. Then $L$ is accepted by the complete deterministic automaton $\calA(M) = (M,\delta,1,X)$, where for $m\in M$ and $a\in A$,
$$\delta(m,a)=m\,\phi(a).$$
Since $M$ is finite, $L$ is rational.
\eop

\begin{remark}\label{rk: transition monoid}
Observe that if $M$ is a finite monoid and $\calA(M)$ is the automaton defined in the proof of Theorem~\ref{thm:rational-recognizable}, then the transition monoid of $\calA(M)$ is $M$ itself.
\end{remark}

The syntactic monoid plays the same role in this algebraic view of rational languages that the minimal automaton plays in the automaton-theoretic view.  Here we make this precise: We say that a monoid $N$ \emph{divides}\index{division} a monoid $M$, and write $N\prec M$, if there is a submonoid $M'$ of $M$ and a surjective morphism $\phi\colon M'\to N$.  It is easy to see that $\prec$ is a transitive relation on monoids.

\begin{theorem}~\label{thm:syntactic-monoid-minimal}
Let $L\subseteq A^*$.  Then $M$ recognizes $L$ if and only if $M(L)\prec M$.
\end{theorem}

\begin{proof}
First suppose $M$ recognizes $L$, so that there is a morphism $\phi\colon A^*\to M$ such that $L=\phi^{-1}(X)$ for some $X\subseteq M$.  We claim that if $\phi(u)=\phi(v)$, then $u\cong_L v$.  To see this, suppose that $xuy\in L$ for some $x,y\in A^*$. Then $\phi(xuy)\in X$, and since $\phi(u)=\phi(v)$, we have $\phi(xvy)\in X$, so that $xvy\in L$.  By the same argument, if $xvy\in L$ then $xuy\in L$, so that $u\cong_L v$.

Now let $M'=\phi(A^*)$. We define a map $\psi\colon M'\to M(L)$ by $\psi(\phi(u))=\mu_L(u)$.  By the remark just made, $\psi$ is well-defined, since the value of $\psi$ only depends on $\phi(u)$ and not on $u$. Moreover $\psi$ is clearly a morphism, and it is surjective because $\mu_L$ is, so $M(L)\prec M$.

Conversely, suppose $M(L)\prec M$, so that there is a morphism $\psi$ from a submonoid $M'$ of $M$ onto $M(L)$.  For $a\in A$, we set $\phi(a)$ to be any $m\in M'$ for which $\psi(m)=\mu_L(a)$.  This can be extended to a unique morphism $\phi\colon A^*\to M$ such that $\mu_L=\psi\circ\phi$. Let $X=\phi(L)$.  If $\phi(u)\in X$ then  $\phi(u)=\phi(v)$ for some $v\in L$, and thus $\mu_L(u)=\mu_L(v)$, so $u\cong_L v$. Since, as noted in the proof of Theorem~\ref{thm:rational-recognizable}, $L$ is a union of $\cong_L$-classes, this implies $u\in L$, so that $L=\phi^{-1}(X)$, and thus $M$ recognizes $L$.
\eop

\begin{example}\label{ex:syntabstar}
Consider the transition monoid of the automaton $\calA$ in Figure~\ref{fig: with a group}.  
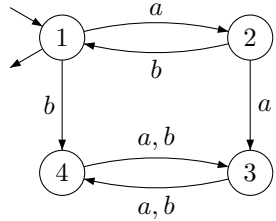
\begin{figure}[t]
\begin{center}
\begin{picture}(30,24)(0,0)
\node[iangle=150,fangle=-150,Nmarks=if,Nw=6.0,Nh=6.0](A)(0,18){$1$}
\node[Nw=6.0,Nh=6.0](B)(25,18){$2$}
\node[Nw=6.0,Nh=6.0](C)(25,0){$3$}
\node[Nw=6.0,Nh=6.0](D)(0,0){$4$}

\drawedge[curvedepth=2](A,B){$a$}
\drawedge[curvedepth=2](B,A){$b$}
\drawedge[curvedepth=2](D,C){$a,b$}
\drawedge[curvedepth=2](C,D){$a,b$}
\drawedge[ELside=r](A,D){$b$}
\drawedge(B,C){$a$}
\end{picture}
\end{center}
\caption{An automaton, whose transition monoid contains a non-trivial group}
\label{fig: with a group}
\end{figure}
We can fairly easily determine the elements of this monoid without doing an exhaustive tabulation:  First, if a word $w$ has even length, then it maps $\{1,3\}$ into $\{1,3\}$, and $\{2,4\}$ into $\{2,4\}$,  while if $w$ has odd length, then it interchanges these two sets.  Second, if $w$ contains $aa$ or $bb$ as a factor, then the image of $f_w$ is contained in $\{3,4\}$.  Finally, if the letters of $w$ alternate, then $f_w$ maps either 1 or 2 to $\{1,2\}$, but not both, depending on whether the first letter of $w$ is $a$ or $b$. We thus get these elements:
\begin{align*}
1 =  f_\epsilon  = &\ (\begin{array}{cccc}1&2&3&4\end{array})\qquad
\gamma =  f_a  =(\begin{array}{cccc}2&3&4&3\end{array})\qquad
 \delta =  f_b  =(\begin{array}{cccc}4&1&4&3\end{array})\\
 &\gamma\delta =  f_{ab}  =(\begin{array}{cccc}1&4&3&4\end{array})\qquad \delta\gamma =  f_{ba}  =(\begin{array}{cccc}3&2&3&4\end{array})\\
&\gamma^2 =  f_{aa}  =(\begin{array}{cccc}3&4&3&4\end{array})\qquad \gamma^3 =  f_{aaa}  =(\begin{array}{cccc}4&3&4&3\end{array})
\end{align*}
Observe that $\{\gamma^2,\gamma^3\}$ forms a group, permuting the states $3$ and $4$.  This automaton accepts the language $(ab)^*$.  In algebraic terms, the morphism $\phi\colon w\mapsto f_w$ from $A^*$ into $M(\calA)$ recognizes this language with $(ab)^*=\phi^{-1}(X)$, where 
$$X=\{f\in M(\calA) \mid \hbox{$f$ maps state 1 to itself }\}.$$
The states 3 and 4 are equivalent, and the minimal automaton of $L$ is obtained by merging these states: it is the automaton examined in  Example~\ref{ex:monoid-abstar} (with 1 as initial and final state), where we computed its transition monoid, namely $M(L)$.  According to Theorem~\ref{thm:syntactic-monoid-minimal}, $M(L)\prec M(\calA)$, and, indeed, the map sending 1 to 1, $\gamma,\delta,\gamma\delta,\delta\gamma$ to $\alpha,\beta,\alpha\beta,\beta\alpha$, respectively, and $\gamma^2$ and $\gamma^3$ both to 0, is a morphism from $M(\calA)$ onto $M(L)$.
\end{example}

\begin{example} Let us take the automaton of Example~\ref{ex:fulltm} and specify 1 as both the initial state and the unique accepting state. With these choices, the automaton is the minimal automaton of the language it accepts, since every state is accessible and no two distinct states are equivalent.  This shows that the syntactic monoid of a language accepted by an $n$-state automaton can have as many as $n^n$ elements.

\end{example}

\begin{example}
Not every finite monoid is the syntactic monoid of a rational language.  Consider, for instance, the monoid $M=\{1,\alpha,\beta,\gamma\}$ with multiplication $m_1m_2=m_2$ for $m_2\neq 1$. Suppose $A$ is a finite alphabet and $\phi\colon A^*\to M$ is a morphism.  Let $X\subseteq M$.  We partition $A$ into three subsets, $B$, $C$, and $D$, 
\begin{align*}
B &= \{a\in A \mid \phi(a)=1\} \\
C &= \{a\in A \mid \phi(a)\in X\setminus\{1\}\} \\
D &= A\setminus(B\cup C)
\end{align*}

Then $\phi^{-1}(X)=B^*\cup A^*CB^*$ if $1\in X$ and $\phi^{-1}(X)=A^*CB^*$ otherwise. (Observe that $B$ or $C$ might be empty.)  But then
$L=\phi^{-1}(X)$ is recognized by the submonoid $\{1,\alpha,\beta\}$, using the morphism  that maps $B$ to 1, $C$ to $\alpha$ and $D$ to $\beta$. Thus every language recognized by $M$ is recognized by a strictly smaller monoid, so by Theorem~\ref{thm:syntactic-monoid-minimal}, $M$ cannot be the syntactic monoid of any language.
\end{example}

\section{First-order definable languages}\label{sec: FO SF Ap}

This section is devoted to proving one of the earliest and most important applications of the syntactic monoid: the characterization of the languages definable in $\FO(<)$.

A finite monoid $M$ can contain a nontrivial group, as for example the group $\{\gamma^2,\gamma^3\}$ in the monoid $M(\calA)$ of Example~\ref{ex:syntabstar}.  If there is no nontrivial group in $M$, we say that $M$ is \emph{aperiodic}\index{monoid!aperiodic}.

\begin{lemma}\label{lemma:aperiodic}
Let $M$ be a finite monoid.  Then the following are equivalent:
\begin{enumerate}
\item $M$ is aperiodic.
\item There is an integer $n>0$ such that $m^n=m^{n+1}$ for all $m\in M$.
\end{enumerate}
\end{lemma}

\begin{proof}
Suppose $M$ is aperiodic.  Let $m\in M$, and consider the sequence $1,m,m^2,\ldots$  Since $M$ is finite, if we take $n=|M|$, we have $m^r=m^n$ for some $r<n$.  Take the largest such $r$, and consider the set $G=\{m^k \mid r\le k< n\}$. Observe that for all $g\in G$, $gG=Gg=G$, since 
$$m^{r+t}m^s=m^{r+[{(t+s)\bmod{(n-r)}}]}$$
for all $s,t\geq 0$.  This implies that $G$ is a group, so that $|G|=1$, and thus $r = n-1$ and $m^r=m^{r+1}$.  Conversely, if $M$ is not aperiodic, then $M$ contains a nontrivial group $G$, and an element $g\in G$ different from the identity element $e$ of $G$. Then $g^k=e$ for some $k>1$, so that $g^n\neq g^{n+1}$ for all $n \ge 0$.
\eop

Note that the proof shows that we can choose $n$ in condition {\it (2)} of Lemma~\ref{lemma:aperiodic} to be $|M|-1$.

We say that a language $L\subseteq A^*$ is \emph{star-free}\index{language!star-free} if it can be defined by an extended rational expression without the use of the $*$ operation or morphic images. The Sch\"utzenberger-McNaughton-Papert Theorem offers the following characterization.

\begin{theorem}\label{thm:schutz-mcnaughton}
Let $L\subseteq A^*$ be a rational language.  Then the following are equivalent.
\begin{enumerate}
\item $L$ is star-free.
\item $L$ is definable by a sentence of $\FO(<)$.
\item $L$ is recognized by an aperiodic finite monoid.
\item $M(L)$ is aperiodic.
\end{enumerate}
\end{theorem}

Before we turn to the proof of this theorem, we give an important corollary, and an example.

\begin{corollary}\label{cor: decidability FO}
It is decidable whether a rational language (given by a rational expression or an accepting automaton) is definable by a sentence of first-order logic.
\end{corollary}

\begin{proof} As we have seen, we can compute $\calA_{\rm min}(L)$ from any automaton or expression for $L$, and thence compute the multiplication table of $M=M(L)$.  We can then test for all $m\in M$ whether $m^{|M|-1}=m^{|M|}$, and thus, by Lemma~\ref{lemma:aperiodic} determine whether $M(L)$ is aperiodic.  By Theorem~\ref{thm:schutz-mcnaughton}, this decides whether $L$ is first-order definable.
\eop

In fact, the proof of Theorem~\ref{thm:schutz-mcnaughton} will show that if $M(L)$ is aperiodic, then we can effectively construct both a star-free expression and a first-order sentence for $L$ from an automaton that recognizes $L$.

\begin{example}
Let $L=(ab)^*$.  We computed $M(L)$ in Example~\ref{ex:monoid-abstar}.  We have $\alpha^2=\beta^2=0=\alpha^3=\beta^3$, and $(\alpha\beta)^2=\alpha\beta$, $(\beta\alpha)^2=\beta\alpha$, so by Lemma~\ref{lemma:aperiodic}, $M(L)$ is aperiodic.  Theorem~\ref{thm:schutz-mcnaughton} says that $L$ is definable by a star-free extended rational expression, and also by a sentence of $\FO(<)$. Let us exhibit such expressions.

First, note that membership of a word $w$ in $L$ is equivalent to saying that $w$ contains no occurrence of either $aa$ or $bb$ as a factor, and that the first letter of $w$ (if there is one) is $a$, and the last letter is $b$.  We thus have
$$L = \{\epsilon\}\cup(aA^*\cap A^*b \cap \overline{A^*(aa\cup bb)A^*}).$$
witnessing the fact that $L$ is star-free (note that $A^*$ is star-free, since $A^*=\overline{\emptyset}$).

To obtain a first-order sentence defining $L$, we use the same characterization of words in $L$. We say there is no occurrence of $aa$ as a factor using the following sentence:
$$\neg\exists x\exists y(R_ax \wedge R_ay \wedge S(x,y)).$$
This uses the successor predicate $S$, but as we noted earlier, $S$ can be expressed in $\FO(<)$.  We can likewise write a sentence saying that there is no occurrence of $bb$. An $\FO$-sentence stating that the first letter of a word is $a$ was given in Example~\ref{ex logical aA^*}. A similar sentence can be formed to say that the last letter is $b$.  Note that all these sentences are satisfied by the empty word as well, so that the conjunction of the four sentences defines the language $(ab)^*$.

This language is also recognized by the first monoid that we exhibited in Example~\ref{ex:syntabstar}, which is not aperiodic.   This in no way contradicts Theorem~\ref{thm:schutz-mcnaughton}, which only says that {\it some} aperiodic monoid recognizes $L$.
 \end{example}
 
\begin{remark}\label{addl rk 4}
The decision procedure outlined in the proof of Corollary~\ref{cor: decidability FO} may take exponential time in the size of an automaton accepting $L$, since it involves computing the syntactic monoid of $L$ (see Example~\ref{ex:fulltm}). While this procedure may be improved, this decision problem is intrinsically difficult. In fact, it is known to be PSPACE-complete (Cho and Huynh \cite{Cho&Huynh:1991}).
\end{remark}

 We  now turn to the proof of Theorem~\ref{thm:schutz-mcnaughton}.  We will show $(4) \Leftrightarrow (3) \Rightarrow (1) \Rightarrow (2) \Rightarrow  (4)$. By Theorem~\ref{thm:syntactic-monoid-minimal}, every language is recognized by its syntactic monoid. Also every divisor of an aperiodic monoid is aperiodic, since the property $m^n=m^{n+1}$ for all elements $m$ in a monoid is inherited by morphic images and submonoids.  Thus (3) and (4) are equivalent.
 
The most difficult part of the proof is $(3) \Rightarrow (1) $.  To prove this, we suppose $L\subseteq A^*$ is recognized by a finite aperiodic monoid.  This is equivalent to $L$ being accepted by a complete deterministic automaton $\calA = (Q,\delta,i,F)$ whose transition monoid is aperiodic (see the proof of Theorem~\ref{thm:rational-recognizable} and Remark~\ref{rk: transition monoid}).  We will show that for all $q,q'\in Q$, the set $L_{q,q'}^{\calA}=\{w \mid qf_w^{\calA}=q'\}$ is a star-free language. Since  $L$ is a finite union of such languages, $L$ is star-free.
    
The proof is by induction on the pair $(|Q|,|A|)$: the induction hypothesis is that the claim holds for all automata with a strictly smaller state set, or with the same size state set and a strictly smaller input alphabet.  In the case $|Q|=1$, $L$ is either $A^*$ or $\emptyset$, which are star-free. In the case $|A|=1$, so that $A=\{a\}$, aperiodicity implies that  $L$ is a finite union of singleton sets $\{a^k\}$, possibly together with the language $a^ra^*$, where $r=|Q|-1$,  which is also star-free, since $a^*=\overline{\emptyset}$.
    
 We thus assume both $|Q|>1$ and $|A|>1$. First suppose that for every $a\in A$, $Qf_a^{\calA}=Q$, so that $f_a^{\calA}$ is a permutation of $Q$. Aperiodicity implies $(f_a^{\calA})^r=(f_a^{\calA})^{r+1}$ for some $r$, and thus $f_a^{\calA}$ is the identity map on $Q$.  Consequently $f_w^{\calA}$ is the identity map for all $w\in A^*$, and thus the claim holds trivially.  We can therefore assume that there is some $a\in A$ such that 
$$Qf_a^{\calA} = Q' \subsetneq Q.$$
We now define two new automata $\calB$ and $\calC$. Automaton $\calB$ has state set $Q$ and next-state function $\delta\big\vert_{Q\times B}$, where $B=A \setminus \{a\}$.  We need not define initial and final states for $\calB$, because we are only interested in the state transitions $f_w^{\calB}$. Automaton $\calC$ has state set $Q'$, input alphabet 
$$C=\{(f_w^{\calB},a) \mid w\in B^*, a\in A\},$$
and next-state function
$$\delta'\colon (q,(f_w^{\calB},a))\longmapsto q\cdot f_{wa}^{\calA}.$$
This makes sense, because $Qf_a^{\calA}=Q'$ and
because $f_w^\mathcal{B}=f_{w'}^\mathcal{B}$ implies that 
$f_{wa}^\mathcal{A}=f_{w'a}^\mathcal{A}$.  The inductive hypothesis applies to both $\calB$ and $\calC$. (The  transition monoids of these automata inherit the aperiodicity of $\calA$, because every transition in them is the restriction of a transition in $\calA$.) 

A word in $L_{q,q'}^{\calA}$ can contain either no occurrences of $a$, a single occurrence of $a$, or two or more occurrences of $a$.  We can accordingly write $L_{q,q'}^{\calA}$ as a finite union of sets of the form
$$L_{q,q'}^{\calB}, L_{q,p}^{\calB}aL_{p',q'}^{\calB},L_{q,p}^{\calB}aT_{p',q'' }L_{q'' ,q'}^{\calB},$$
where $p\in Q$, $p'=p\cdot f_a^{\calA}\in Q'$, and $T_{p,q''}=L_{p,q''}^{\calA}\cap A^*a$.

 By the inductive hypothesis all the sets of the form $L_{s,t}^{\calB}$ are star-free, so it remains to show that $T_{p',q''}$ is a star-free language.  We can factor any $w\in A^*a$ uniquely as
$$w=v_1a\cdots v_ka,$$
where $v_1,\ldots, v_k\in B^*$.  Let us associate to $w$ the word
$$w_C=c_1\cdots c_k\in C^*,$$
where $c_j=(f_{v_j}^{\calB},a)\in C$.  By the inductive hypothesis, the language $L_{p',q''}^{\calC}$ is star-free.  So we need to show that if $R\subseteq C^*$ is star-free, then $\Psi(R)=\{w\in A^*a \mid w_C\in R\}$ is also star-free, since $T_{p'q''}=\Psi(L_{p',q''}^{\calC})$. 
 It is thus enough to show
 \begin{itemize}
 \item[\textit{(i)}] If $c\in C$, then $\Psi(\{c\})$ is star-free.
 \item[\textit{(ii)}] If $\Psi(R)$ is star-free, then $\Psi(C^*\setminus R)$ is star-free.
 \item[\textit{(iii)}] If $\Psi(R_1),\Psi(R_2)$ are star-free, then $\Psi(R_1\cup R_2)$ is star-free.
 \item[\textit{(iv)}] If $\Psi(R_1),\Psi(R_2)$ are star-free, then $\Psi(R_1 R_2)$ is star-free.
\end{itemize}
 For \textit{(i)}, note that $\Psi(\{c\})=Sa$, where $S=\{v\in B^* \mid c=(f_v^{\calB},a)\}$.  Since $S$ is a boolean combination of languages of the form $L_{p,p'}^{\calB}$, $Sa$ is star-free.
 For the other assertions, we clearly have $\Psi(C^*\setminus R)=A^*a\cap (A^*\setminus \Psi(R))$,  $\Psi(R_1\cup R_2)=\Psi(R_1)\cup \Psi(R_2)$, and $\Psi(R_1R_2)=\Psi(R_1)\Psi(R_2)$. This completes the proof that $\textit{(3)} \Rightarrow \textit{(1)}$.

To prove $(1) \Rightarrow (2)$, we need to show that every star-free language is first-order definable.  Since the singleton sets $\{a\}$ for $a\in A$ are clearly first-order definable, and since the boolean operations are part of first-order logic, this reduces to showing that if $L_1,L_2\subseteq A^*$ are first-order definable, then so is $L_1L_2$.  To do this, we introduce the notion of \emph{relativizing}\index{relativization} a first-order sentence.  Let $\phi$ be a sentence of $\FO(<)$ and $x$ a variable symbol that does not occur in $\phi$.  We define a formula $\phi_{<x}$ with one free variable with the following property:  Let $\nu$ be an interpretation mapping $x$ to $i\in \Dom(u)$, and let $v$ be   the prefix $v$ of $u$ with domain $\{0,\ldots,i-1\}$. Then $u,\nu\models\phi_{<x}$ if and only if $v\models\phi$.  To construct $\phi_{<x}$, we simply work from the outermost quantifier of $\phi$ inward, replacing each quantified subformula $\exists y\,\alpha$ by $\exists y\,((y<x)\wedge\alpha)$. We define $\phi_{>x}$ and $\phi_{\leq x}$ analogously.

Now suppose $\phi,\psi$ are first-order sentences defining $L_1$ and $L_2$, respectively. Let $x$ be a variable symbol that does not occur in $\phi$ or $\psi$. We have
$L_1L_2$ defined by the sentence 
\begin{align*}
\exists x\ (\phi_{\leq x}\wedge\psi_{>x})\qquad&\textrm{if $\epsilon \notin L_1$,}\\
\exists x\ (\phi_{\leq x}\wedge\psi_{>x})\vee \psi\qquad&\textrm{if $\epsilon \in L_1$.}
\end{align*}

To prove $(2)\Rightarrow (4)$, we need to show that the syntactic monoid of every first-order definable language in $A^*$ is aperiodic. We will proceed as in Section~\ref{sec: formulas 2 expressions}, and treat a first-order formula with free variables contained in $\{x_1,\ldots, x_p\}$ as defining a language over the extended alphabet $B_p=A\times\{0,1\}^p$.  We will show by induction on the quantifier depth that every first-order definable language $L\subseteq B_p^*$ in this extended sense has an aperiodic syntactic monoid.  More precisely, we will show that for each such $L$ there exists an integer $q>0$ such that for all $v\in B_p^*$, $v^q\cong_Lv^{q+1}$.  By Lemma~\ref{lemma:aperiodic}, this implies aperiodicity.

First suppose $L$ is defined by one of the atomic formulas $x_1<x_2$ or $R_ax_1$.  Let $u,v,w\in B_p^*$.  If $v$ has a letter with a 1 in one of its last $p$ components, then neither $uv^2w$ nor $uv^3w$ can be in $L$, since only one letter of a word in $L$ can have a 1 in a given component.  If $v$ has no such letter,  then membership of $uvw$ in $L$ is witnessed by the relative positions and values of letters in $u$ and $w$, so that $uvw\in L$ if and only if $uv^2w\in L$.  Thus in all cases, we have $uv^2w\in L$ if and only if $uv^3w\in L$, so that $v^2\cong_L v^3$.

Now suppose the claim is true for $L_1, L_2\subseteq B_p^*$ defined by formulas $\phi_1,\phi_2$, and suppose $L$ is defined by $\phi_1\vee\phi_2$.  We have, by assumption, $v^q\cong_{L_1} v^{q+1}$, and $v^q\cong_{L_2} v^{q+1}$, for some $q>0$.  (The exponents for these two languages are, {\it a priori}, different, but we can then choose $q$ to be the maximum of the two exponents.) Now $\phi_1\vee\phi_2$ defines $L_1\cup L_2$, and we have directly $uv^qw\in L_1\cup L_2$ if and only if $uv^{q+1}v\in L_1\cup L_2$.  

Care must be taken with the negation operator, since it does not exactly correspond to the boolean complement. We can assume that the exponent $q$ for $L_1$ is at least 2. Let $L_1'$ be the language defined by $\neg\phi_1$. Suppose $uv^qw\in L_1'$.   Then $uv^qw\notin L_1$, and thus $uv^{q+1}w\notin L_1$.  Further  $v$ cannot contain a 1 in the last 
$p$ components of any of its positions, so  $uv^{q+1}w$ has exactly one occurrence of 1 in each of the last $p$ positions, and thus is in $L_1'$.   The same argument shows that if $uv^{q+1}w\in L_1'$,  then so is $uv^qw$.  Thus $v^q\cong_{L_1'} v^{q+1}$.

So now let $K\subseteq B_{p-1}^*$ be the language defined by $\exists x_p\phi_1$. Let $v\in B_{p-1}^*$. We will show $v^{2q+1}\cong_K v^{2q+2}$. Suppose $uv^{2q+1}w\in K$.  Let us extend each letter in this word by adding a $p^{th}$ component with 0.  We will still denote the resulting word as $uv^{2q+1}w$. Since $K$ is defined by $\exists x_p\phi_1$,  we can switch the $p^{th}$ component of some letter to obtain a word $z\in B_p^*$ such that $z\in L_1$. Now, wherever the position in which we switched the $p^{th}$ component is located, at least $q$ consecutive occurrences of $v$ will be left intact.  We thus find that $z$ can be written in the form $xv^qy$, for some $x,y\in B_p^*$. (The extreme case is when the position is within the middle occurrence of $v$, in which case we get two factors of the form $v^q$.) Thus $xv^{q+1}y\in L_1$.  If we now switch the changed 1 back to 0, we find $uv^{2q+2}w\in K$. The identical argument shows $uv^{2q+2}w\in K$ implies $uv^{2q+1}w\in K$.  Thus $v^{2q+1}\cong_K v^{2q+2}$, as claimed.

\begin{remark}
Interesting presentations of proofs of all or part of Theorem~\ref{thm:schutz-mcnaughton} can be found, for instance, in the work of Perrin~\cite{Perrin1990handbook}, Straubing~\cite{Straubing:1994} and Diekert and Gastin~\cite{DiekertGastin2008siwt}.
\end{remark}

\bibliographystyle{plain}



\printindex

\end{document}